\documentclass[%
 aip,
 jmp,%
 amsmath,amssymb,
 reprint,%
]{revtex4-1}

\usepackage{graphicx}
\usepackage[percent]{overpic}
\usepackage{dcolumn}
\usepackage{bm}
\usepackage{color}
\usepackage{hyperref}

\newcommand{\sol}{s}  
\newcommand{\liq}{l}  
\newcommand{\vap}{v}  
\renewcommand{\sl}{{sl}}  
\newcommand{\lv}{{lv}}  
\newcommand{\sv}{{sv}}  
\newcommand{\change}[1]{ {#1} }

\begin{document}

\preprint{}


\title[]{The Key Physics of Ice Premelting}

\author{Luis G. MacDowell}
\email{lgmac@quim.ucm.es}
\affiliation{Departamento de Qu{\'i}mica F{\'i}sica,
	Facultad de Ciencias Qu{\'i}micas, Universidad Complutense de Madrid, Spain.
}%

\date{\today}

\begin{abstract}
	A disordered quasi-liquid layer of water is thought to cover the ice surface, \change{but many issues such as}  its onset temperature, its thickness, or its actual relation to bulk liquid water have been a matter of unsettled controversy for more than a century.  In this perspective article, current computer simulations and experimental results are discussed under the light of a suitable theoretical framework. It is found that using a combination of wetting physics, the theory of intermolecular forces, statistical mechanics and out of equilibrium physics  a large number of conflicting results can be reconciled and collected into a consistent  description of the ice surface. This helps understand the crucial role of surface properties in a range of important applications, from the enigmatic structure of snow crystals to the  slipperiness of ice.	
\end{abstract}

\pacs{Valid PACS appear here}

\maketitle


\section{Introduction}

\begin{figure}
	\includegraphics[width=0.5\textwidth,trim=0 0 17cm 0,clip]{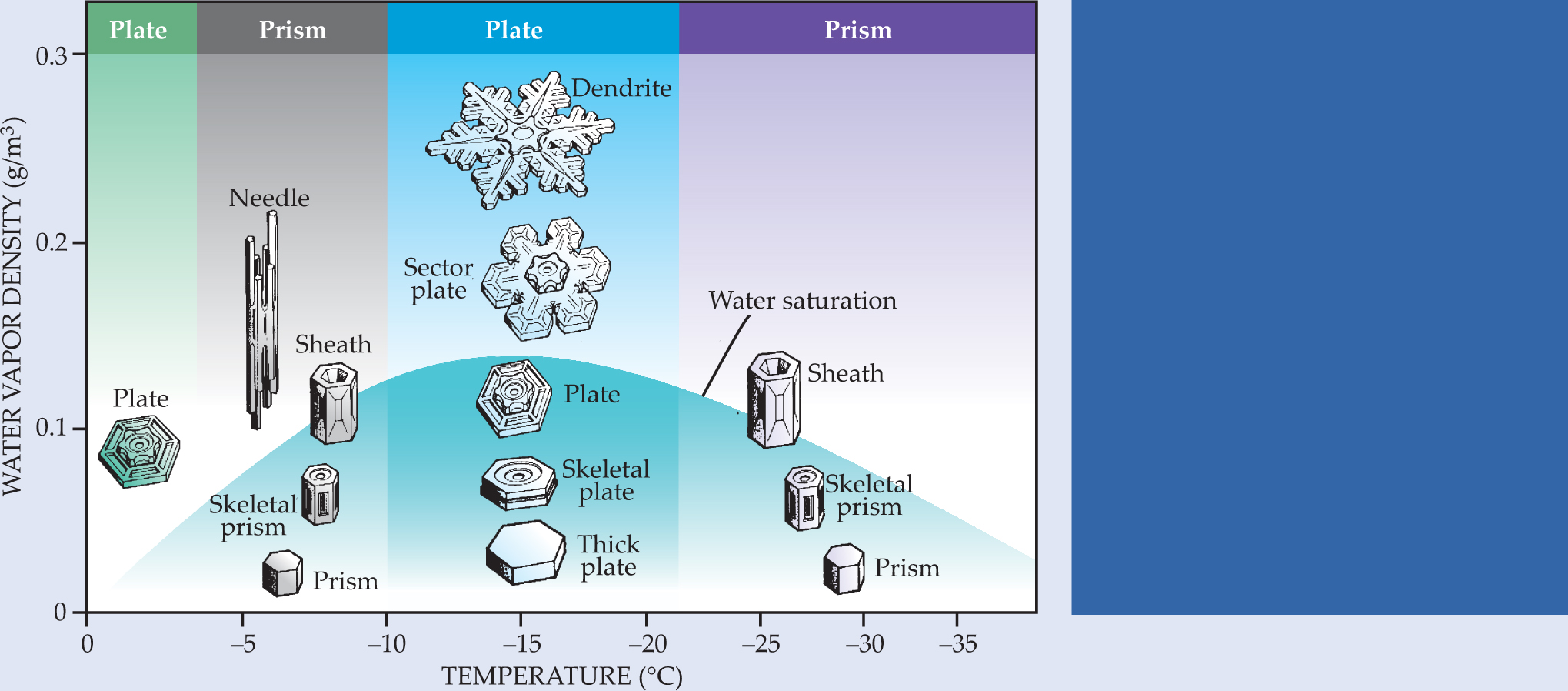}
	\caption{Diagram of snow crystal morphology, a descendant of the early
	   Nakaya diagram due to Kobayashi.\cite{kobayashi57} The figure displays
	   the habit of ice crystals grown in the atmosphere as a function of
	   temperature and excess water vapor density over ice saturation. The
	   origin of the complex shapes that appear as saturation increases are
	   relatively well understood. The reason why simple hexagonal prisms
	   grown at low vapor saturation change from plates, to columns and back
	   again to plates as temperature decreases has remained a mystery ever
	   since Nakaya first published his results. Reproduced from \href{https://doi.org/10.1063/1.2825081}{Furukawa and
	   Wettlaufer, Phys. Today} {\bf 60} 70 (2007),\cite{furukawa07} with the
	   permission of \change{AIP publishing}.
		\label{fig:nakaya}
		 }
\end{figure}

On the 12th of March 1936 a team of Japanese researchers led by Ukichiro Nakaya succeeded in creating
the first ever artificial snow crystal in a snow chamber under controlled
conditions.\cite{nakaya54}  After several years
of further research, the results were summarized and collected in the diagram
that now bears his name.\cite{nakaya54,kobayashi57,frank82,furukawa07,libbrecht22} The
Nakaya diagram  maps crystal habits as  a function of  temperature and
humidity of the surrounding atmosphere (Figure~\ref{fig:nakaya}). The complex
structures that form as supersaturation is increased can be relatively well
understood under the framework of crystal growth science.\cite{frank82,nelson01,libbrecht22} 
However, the surprising transformation of solid hexagonal prisms  from plate
like to columnar like and back to plate like as temperature decreases has remained a mystery 
ever since.\cite{frank82,furukawa07,libbrecht22}

Nakaya understood that the only way to interpret the experimental observations was to
assume the surface structure of ice must be highly complex. For this
reason, he revisited an old hypothesis by Faraday, who first conjectured that
ice below its melting point  bares a microscopically thin liquid layer on its
surface.\cite{faraday50} Based on this idea, he performed experiments, later
improved by Hosler and Goldshlak, to measure the adhesion of ice
beads.\cite{nakaya54b,hosler57} The teams concluded that the adhesion  and
its dependence on humidity  was likely to result from the
premelting layer conjectured by Faraday.  However, due to the lack of adequate
experimental tools, the hypothesis remained unconfirmed. Delightful reports on
the ingenuity and perspicacity of early researchers to probe the ice
surface for a premelting film using rudimentary experimental apparatus may be found in 
Ref.\cite{weyl51,jellinek67,rosenberg05}

Two decades later, Kuroda and Lacmann borrowed an original idea from Stransky on 
premelting mediated crystal growth,\cite{lacmann72} and formulated a consistent theory 
of ice growth from the vapor.
Their working hypothesis was that the basal and prism faces of ice undergo a
sequence of surface phase transitions. Switching from one surface phase to
another would cause abrupt shifts in the relative growth rates of the basal and
prism faces, thereby accounting for the puzzling plate-to-column transformations
depicted in the Nakaya diagram.\cite{kuroda82,kuroda83} 

Kuroda went on to conduct ellipsometry and x-ray diffraction studies that were
among the first few direct evidences for the presence of a quasi-liquid layer of
melt water on the ice surface.\cite{furukawa87,kouchi87} Following these
studies, the field experienced a surge of renewed activity, with numerous
measurements and estimates of the quasi-liquid layer thickness carried out using
a variety of modern spectroscopic and mechanical techniques.\cite{kouchi87,elbaum93,lied94,dosch95,dosch96,wei01,doppenschmidt00,bluhm02,sadtchenko02,constantin18,mitsui19} 

Within a decade, these experimental measurements confirmed what had remained a plausible hypothesis for more than a hundred years.

Paradoxically, the experimental confirmation of the premelting phenomenon in ice did not settle the issue altogether. Estimates of the precise thickness of the quasi-liquid layer varied by orders of magnitude, according to most reviews.\cite{petrenko94,dash06,bartels-rausch14,slater19}
In this context, the discussion shifted to whether the quasi-liquid layer
maintains a finite thickness up to the triple point, or whether it diverges and
becomes infinitely thick. In the language of adsorption science, this corresponds to asking 
whether a complete surface melting transition exists (the complete wetting of
water on ice), or whether surface melting remains incomplete at three phase coexistence
(the partial wetting of water on ice).\cite{schick90}

Regardless of actual measurements of the quasi-liquid layer thickness as
a function of temperature, experimental observations dating back to the late
1960s showed that water does not completely wet the ice surface. Early observations
of water droplets formed on ice,\cite{knight67,ketcham69,knight71} were confirmed several years later,\cite{elbaum91,elbaum93,gonda99} and have been reported during the last decade with high quality imaging.\cite{sazaki12,sazaki13,asakawa15,asakawa16,murata16,demmenie25,sarlin25}
Surprisingly, this implies that the surface of ice is mildly hydrophobic, or at least not completely hydrophilic.

Interestingly, once the presence of liquid water droplets on the ice surface was confirmed by direct optical microscopy imaging, Faraday's original hypothesis came under renewed scrutiny.\cite{knight96} More recently, Sazaki and collaborators--at the Institute for Low Temperature Science, founded by Nakaya and later home to Kuroda--have raised serious concerns about this hypothesis,\cite{sazaki12,asakawa16,murata16,sazaki22}, while  studies on the ice surface using macroscopic probes have lead Bonn et al. to question the significance of ice premelting altogether.\cite{liefferink21,demmenie22,demmenie25}

Of course, ice premelting is thought to have important implications in many
fields other than crystal growth. This includes atmospheric
physics,\cite{bartels-rausch14} glaciology, soil science,\cite{dash06} and tribology.\cite{lever21} However,  disputes and controversies  on the role of premelting films extend into nearly all these topics.\cite{weyl51,jellinek67,rowlinson71,rosenberg05,dash06,bartels-rausch14,slater19,lever21b} In this introduction, I have highlighted its relevance to crystal growth science,
 as seeking for an explanation of the  Nakaya diagram has proved to be an exceptional motivation to delve into a variety of important topics in chemical physics, including
 wetting physics, capillary wave and  crystal growth theory, renormalization theory, intermolecular forces, liquid-state theory, and out of equilibrium physics.
 
 In this perspective article I discuss how a selected blend of concepts from
these wide range of fields together with the help of computer
simulations,\cite{benet16,benet19,macdowell19,llombart19,llombart20,llombart20b,sibley21,luengo21,luengo22b,baran22,baran24,baran24b,baran25}
allows to interpret and reconcile a  large body of apparently conflicting empirical evidence into a unified and consistent picture of the structure and dynamics of premelting films.  \change{This includes some of the most relevant puzzles reported in recent reviews:\cite{slater19,nagata19,sazaki22} How can the observation of elementary steps be reconciled with the presence of significant premelting? why do liquid droplets appear well above water-vapor saturation?  And most notably, what is the reason for the alternation of crystal habits in the Nakaya diagram?}
 
Often, this  requires delving into rather intricate and subtle
discussions. The readers less interested in the finer details of wetting physics and crystal growth 
can get an overall impression of the field by going through sections 
\ref{sec:hvT}, \ref{sec:drops},  \ref{sec:noneq}, \ref{sec:ql} and \ref{sec:concl} 
only, while
a read through section \ref{sec:concl} 
may be sufficient to
get a broad picture of the current understanding and future challenges.

\section{Modern empirical evidence of ice premelting}

\label{sec:hvT}

\begin{figure}
	\begin{center}		
		\includegraphics[width=0.5\textwidth]{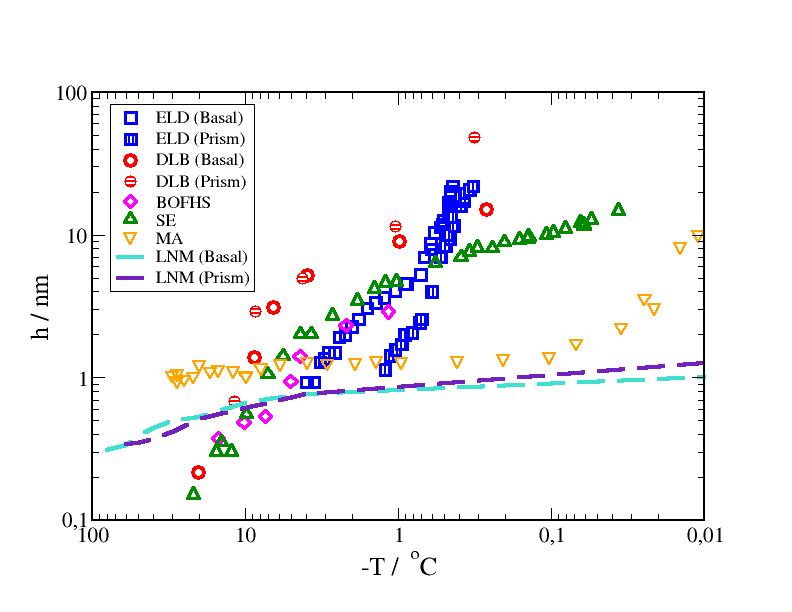}
		\caption{Extent of ice premelting. A selection of all non-invasive
			experimental results since 1990 for temperatures up to -1~$^{\circ}$C lead to
			premelting film thicknesses that agree within an order of magnitude.
			Experimental results from Elbaum et al.\cite{elbaum93} (ELD), Dosch et
			al.\cite{dosch95} (DLB), Bluhm et al.\cite{bluhm02} (BOFHS), Sadtchenko and
			Ewing (SE),\cite{sadtchenko02} and Mitsui and Aoki,\cite{mitsui19} (MA).
			Representative results from computer simulations are provided for the TIP4P/Ice
			model from Llombart et al.\cite{llombart20b} (LNM).
			\label{fig:hvT}}
	\end{center}
\end{figure}

\begin{figure}
	\includegraphics[width=0.5\textwidth]{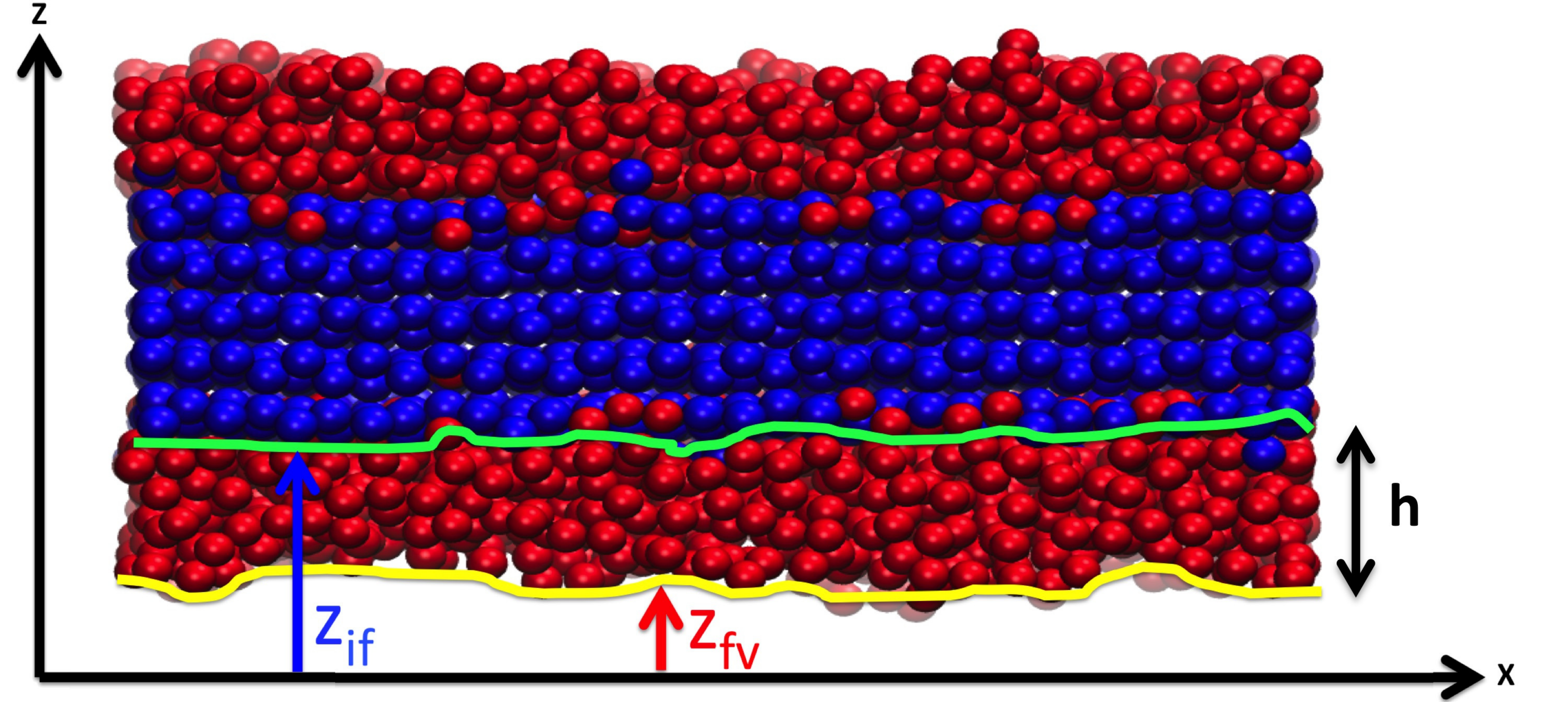}
	\caption{Snapshot of an ice-vapor interface as obtained in computer simulations for the TIP4P/Ice model. Red particles correspond to water molecules in a liquid-like state. Blue molecules are in a solid like state as determined using a suitable order parameter. The interfacial structure may be characterized by two fluctuating surfaces which bound the premelting film from the ice and vapor bulk phases. Reproduced from Llombart et al.\cite{llombart20b} Sci. Adv. {\bf 6} eaay9322 (2020). Distributed under a Creative Commons Attribution License CC-BY.
		\label{fig:snapshot}}
\end{figure}

The experimental measurement of ice premelting thickness has been reviewed in a
large number of
papers.\cite{elbaum93,petrenko94,sadtchenko02,bartels-rausch14,asakawa16,constantin18,slater19}
All of them agree that the data exhibit enormous dispersion, with discrepancies
spanning several orders of magnitude. However, the large extent of scatter can
be reduced significantly by a judicious selection of experimental results.
As noted in Ref.\cite{dash06} there is no reason to expect that the
premelting film thickness should be equal for surface and interfacial
premelting. Indeed, computer simulations show that interfacial premelting
structure is similar to that of surface premelted films only for very hydrophobic
substrates.\cite{baran25} Considering the case of surface premelting
alone reduces significantly the number of available 
measurements.\cite{golecki78,beaglehole80,furukawa87,elbaum93,dosch95,doppenschmidt00,sadtchenko02,constantin18,mitsui19}
However, out of all these studies, the early results by Golecki and Jacard using
proton backscattering are clearly outliers,\cite{golecki78} and can be ruled out. Furthermore,
studies using Atomic Force Microscopy are necessarily invasive, and are prone to
capillary condensation in between the tip and the surface,\cite{pickering18} so they are left out
too. A final selection accounting for all the remaining measurements since 1990,
using widely different non-invasive techniques such as ellipsometry, x-ray
scattering, and infra-red, sum frequency or fluctuation spectroscopy lead to a
rather different picture, with results differing by at most one order of
magnitude for temperatures below about $-1^{\circ}$C (Figure \ref{fig:hvT}). The scatter remains bracketed within 
one order of magnitude even up to temperatures of $-0.1^{\circ}$C if the results of Ref.\cite{dosch96} are discarded.  
All of these experimental measurements confirm that a premelting layer appears on the ice
surface, with thicknesses that increase from the sub-nanometer scale to at most
10~nm in the temperature interval 
between $-20$ and $-1^{\circ}$C. This appears somewhat higher but within
reasonable agreement with computer simulation results for different
water models, which consistently show the formation of a disordered liquid layer
on the ice
surface.\cite{conde08,pfalzgraf11,limmer14,benet16,sanchez17,louden18,qiu18,pickering18,kling18,benet19,llombart19,llombart20,llombart20b}
Indeed, the extent of ice premelting can be quantified in computer simulations using
suitable order parameters that allow to distinguish whether a water molecule is
found in a liquid like or solid like environment (Figure \ref{fig:snapshot}).\cite{lechner08,nguyen15}
Using such parameters, simulation studies report film thicknesses that lie in
the sub-nanometer range up to about  $-1^{\circ}$C below the model's triple point, and
then grow somewhat above the nanometer thickness at higher 
temperatures.\cite{conde08,limmer14,qiu18,pickering18,llombart19,llombart20,llombart20b,berrens22}  

\change{Hower, notice that in strict sense the direct comparison between computer simulations and experiments can only be done for studies probing uncontaminated and atomically smooth ice surfaces with well defined orientation. This requires dedicated preparation of single mono-crystals grown by layer by layer deposition from the vapor.\cite{furukawa87,elbaum93,sazaki12,asakawa15,murata16,libbrecht22} In most  laboratories, measurements have been performed  from rapid vapor deposition,\cite{bluhm02,constantin18} or water freezing,\cite{beaglehole80,sadtchenko02,mitsui19} which produces polycrystalline samples with uncontrolled smoothness unless extra care is exercised.\cite{dosch95} In practice, it seems from Figure \ref{fig:hvT} that the premelting film thickness remains constrained within one order of magnitude irrespective of the smoothness or surface orientation.
Particularly, the current uncertainties in experiments do not seem to settle whether the film thickness is significantly different in basal or prism surfaces, but results from computer simulations suggest this difference is constrained within tenths of a nanometer.\cite{conde08,llombart20,llombart20b,baran24b}}

\section{A toy model of (complete) surface melting}

\label{sec:toymodel}

\begin{figure}[]
	\includegraphics[width=0.4\textwidth]{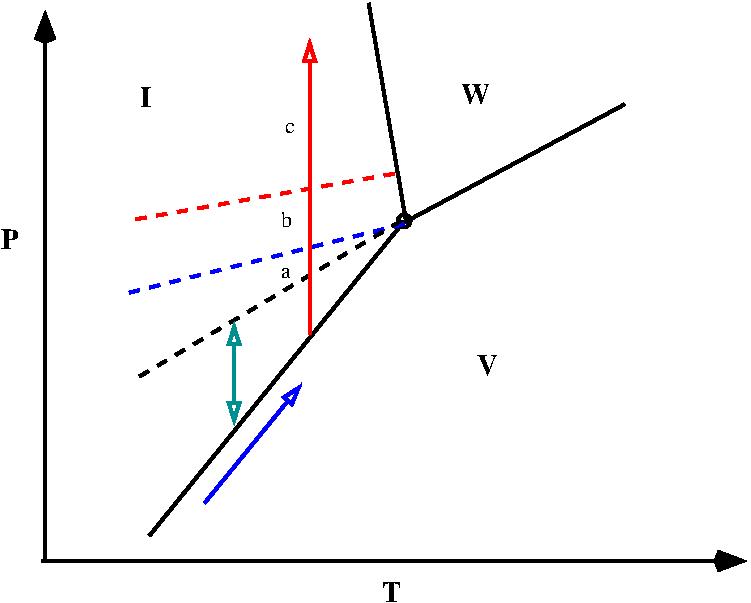}
	\caption[short text]{Phase diagram of water in the neighborhood of the
	triple point. Full black lines illustrate the equilibrium phase
   boundaries, including the {\em sublimation line} separating ice from vapor
and the {\em condensation line}, separating liquid water from vapor. The black
dashed line is the metastable prolongation of the condensation line. Equilibrium
premelting occurs for a path along the sublimation line, as indicated by the
blue arrow. The equilibrium thickness of the premelting film is dictated by the
distance of the sublimation line to the metastable prolongation of the
condensation line, as indicated by the green segment. The red line indicates a
non-equilibrium path, with three different regimes as described in Section IX.
In region {\em a}, below the kinetic condensation line, shown in dashed blue,
ice grows but the premelting film remains in a steady state of constant thickness. In region {\em b}, above the kinetic condensation line but below the kinetic spinodal line shown in dashed red, the premelting film remains in steady state but vapor can condense and form droplets atop. In region {\rm c}, above the kinetic spinodal line, condensation occurs faster than freezing, the premelting film diverges and ice freezes in a wet mode. }
	\label{fig:phase_diagram}
\end{figure}

Despite the empirical evidence discussed in the previous section,
the significance of equilibrium ice premelting has been contested
occasionally,\cite{knight96,asakawa16,demmenie25} so it is now instructive to consider a minimal model of 
the ice/vapor interface approaching three phase coexistence.

To set the scene, let us consider a low temperature bulk solid  in equilibrium with bulk vapor  at two phase coexistence. 
Increasing the temperature while remaining at solid-vapor coexistence, the system moves along the {\em sublimation line}, and approaches the {\em triple point}, where solid, liquid  and vapor phases become equally stable and can coexist simultaneously under the same thermodynamic conditions (Figure \ref{fig:phase_diagram}). 

Premelting, viewed as an equilibrium phenomenon, is the study of how the metastable liquid 
phase ($\liq$) intrudes at the interface between the coexisting bulk solid ($\sol$) and vapor phases
($\vap$) in the vicinity of a metastable liquid-vapor coexistence line \change{(Fig.~\ref{fig:snapshot})}. For many
substances, this can only occur as the system approaches the triple point. For the special case of ice, however, notice the
metastable prolongation of the liquid-vapor coexistence line runs very close to
the sublimation line all the way \change{from the triple point down to very low temperatures}.

In order to gain a microscopic understanding of this problem, let us borrow tools from statistical field theory, and assume there is a suitable {\em order parameter}, $\phi$, which adopts widely different values for the bulk solid  and vapor phases. 

An order parameter can be simply the system's density.  However, in order to characterize the high symmetry of a solid, a more sophisticated order parameter may proof convenient.\cite{lipowsky89,lowen90} For the special case of ice premelting, for example, one can choose some suitable measure of the tetrahedral order,  which allows to distinguish different ice polymorphs from the liquid or vapor phases.\cite{errington01,lechner08,limmer14,nguyen15}

A microscopic characterization of the interfacial properties can now be afforded by introducing an order parameter field, $\phi(z)$, which describes how the order parameter changes continuously from its value in the bulk solid to its value in the bulk vapor along the perpendicular direction to the solid-vapor interface, $z$. 

Within the square gradient approximation, the excess surface free energy \change{of} the interface\change{, $\Delta\omega$,} may be obtained in terms of $\phi(z)$, as a sum of bulk free energies adopted by the system along the interface. 
An additional square gradient term accounts for the cost of creating the inhomogeneity:
\begin{equation}\label{eq:sft}
 \Delta \omega = \int dz \left\{ f(\phi) + \frac{1}{2} C \left( \frac{d \phi}{d z}\right)^2 \right \}
\end{equation}
Here, $f(\phi)$ is the bulk free energy of the system as measured in excess to the value of the bulk coexisting phases, while $C$ is a system's parameter which sets the scale of the surface tension.

This free energy model, already employed by van der Waals, is also known under the name of Cahn-Hilliard theory or  Landau-Ginzburg-Wilson Hamiltonian. In that context, it has proven 
instrumental in understanding the structure of interfaces,\cite{rowlinson82b} nucleation,\cite{cahn59,cahn65} and critical phenomena at two phase coexistence.\cite{binder87,goldenfeld92} Here, 
 one usually introduces a dimensionless order parameter, and assumes the bulk free energy is a simple bi-quadratic potential, $f(\phi)= (1-\phi^2)^2$  with minima $\phi\pm 1$ corresponding to the bulk phases at coexistence. 
 Since Eq.(\ref{eq:sft}) is analogous to a classical Lagrangian, the optimal order parameter field which minimizes the surface free energy satisfies the analog of an energy conservation principle and can be expressed as: 
\begin{equation}\label{eq:EL}
f(\phi) = \frac{1}{2} C \left(\frac{d\phi}{d z}\right)^2
\end{equation}
The solution of this equation for the bi-quadratic potential provides the well known result for the structure of an interface at two phase coexistence:\cite{cahn58,rowlinson82b}
\begin{equation}
  \phi(z) = \tanh(\kappa z)
\end{equation}
where $\kappa$ is an inverse correlation length in the scale of the molecular diameter.

Related models have also been applied previously to the study of premelting, under the assumption that the bulk vapor surface behaves as an inert substrate.\cite{lipowsky82,lipowsky83,limmer14} This maps premelting behavior to a problem that is known in the literature as  short range wetting.\cite{brezin83,schick90} However, this obviates altogether the essence of premelting, which corresponds to the approach of two coexisting phases to a triple point.

\begin{figure*}[t]
	\centering
	\begin{minipage}[b]{0.45\textwidth}
		\begin{overpic}[width=\textwidth]{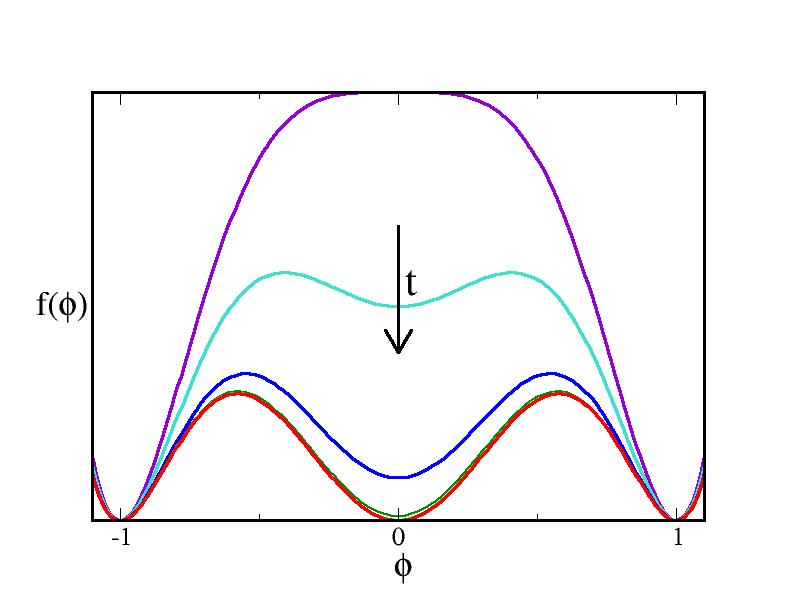}
			\put(15,60){\textbf{(a)}} 
		\end{overpic}
	\end{minipage}
	\hfill
	\begin{minipage}[b]{0.45\textwidth}
		\begin{overpic}[width=\textwidth]{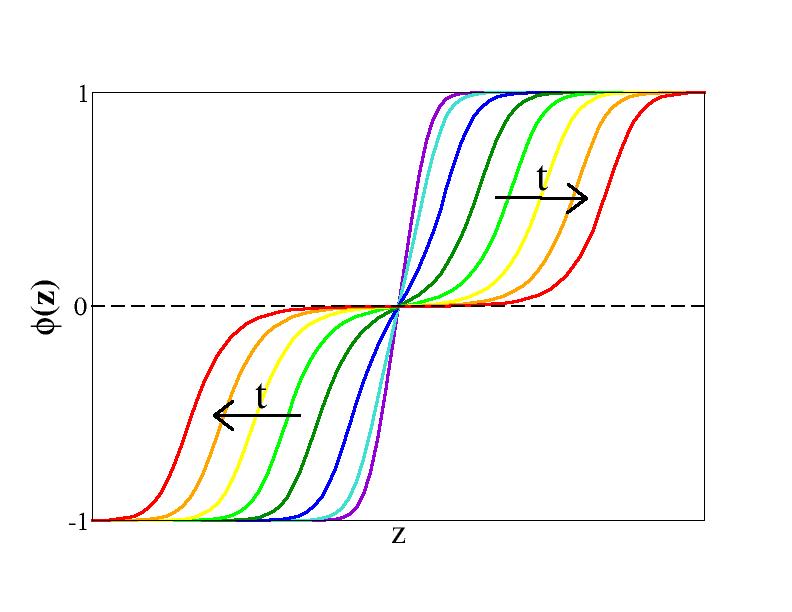}
			\put(15,60){\textbf{(b)}} 
		\end{overpic}
	\end{minipage}
	\caption{Toy model of surface premelting. (a) Free energy of a two-phase system approaching the triple point. At high under-cooling, the free energy exhibits two minima of equal height, corresponding to coexistence of the two phases. Along the sublimation line, a metastable minimum gradually emerges as under-cooling decreases. At the triple point, all three minima reach the same free energy. (b) Corresponding order parameter profiles, showing the gradual enrichment of the metastable phase as under-cooling decreases. Lines from violet to red indicate system behavior in order of decreasing under-cooling, as highlighted by the black arrows.}
	\label{fig:toy}
\end{figure*}

A simple way to remedy this deficiency is to modify the bi-quadratic potential to account for the intrusion of a third phase with an order parameter in between those of the bulk solid and liquid phases, i.e. $f(\phi)= (1-\phi^2)^2 ( t + \phi^2)$, where $t\propto 1-T/T_t$ is a measure of the under-cooling \change{(Figure \ref{fig:toy}-a)}. The Euler-Lagrange equation, Eq.(\ref{eq:EL}) can be solved exactly for this potential, yielding:\cite{lipowsky83}
\begin{equation}\label{eq:phiz}
   \phi(z) = \frac{\sqrt{t} \tanh(\kappa z)}{\sqrt{t + 1 - \tanh^2(\kappa z)}}
\end{equation} 
This result nicely describes the essence of premelting. For deep under-cooling,
$t$ is much larger than $1-\tanh^2(\kappa z)$ for all $z$, and
Eq.(\ref{eq:phiz}) becomes virtually equal to the sharp $\tanh(\kappa z)$
interfacial profile that is typical for two phases at coexistence. When the
system approaches the triple point, however, $t\to 0$, the order parameter
profile adopts a value $\phi(z) \approx \sqrt{t}$ close to zero as long as
$1-\tanh^2(\kappa z)$ in the denominator remains larger than $t$. This occurs in
an increasingly larger range of $z$ values as $t$ vanishes. Accordingly, the
model predicts a gradual increase of the liquid phase in between the solid and
the vapor phases on approaching the triple point (See Figure \ref{fig:toy}\change{-b}).

In order to understand how   this premelting film thickens, we define a film thickness as the distance between the two points where $\phi(z)=\pm 1/2$. Plugging this condition into the profile Eq.(\ref{eq:phiz}), shows that the film thickness presents a logarithmic divergence:
\begin{equation}\label{eq:lndivg}
   h = -\kappa^{-1} \ln t
\end{equation}
Whence, it is predicted that a bulk liquid phase will intrude between the solid and vapor phases at the triple point. This is a case of a {\em complete surface melting} transition, that is akin to the complete wetting transition of a liquid at the solid-vapor interface. The name serves here to emphasize that the premelting film diverges at the triple point, and distinguishes the alternative outcome of {\em incomplete} surface melting, which corresponds to the situation where the premelting film remains finite at the triple point. 
A key concept is to recognize that the complete surface melting of the solid, viewed as an {\em equilibrium} phenomenon, can only occur {\em exactly at the triple point}. Indeed, this is the only thermodynamic state where  three equilibrium macroscopic bulk phases can coexist simultaneously.

To gain further insight into the problem, it is convenient to express the excess surface free energy of Eq.(\ref{eq:sft}) implied by the profile $\phi(z)$ of Eq.(\ref{eq:phiz}) in terms of the film thickness. Using Eq.(\ref{eq:EL}) together with Eq.(\ref{eq:lndivg}) this gives exactly to leading order:
\begin{equation}\label{eq:sfew}
 \Delta \omega(h) = \gamma_{\sl} + \gamma_{\lv} + B e^{-\kappa_t h} - \Delta p h
\end{equation}
where $\gamma_{\sl}$ and $\gamma_{\lv}$ are the solid-liquid and liquid-vapor surface tensions, $B$ is a positive constant in the scale of the surface tension, $\kappa_t$ is the correlation length at the triple point and $\Delta p\propto t$ is the pressure difference between the metastable liquid phase and the coexisting bulk solid or vapor phases.

Another key concept is to acknowledge that $\Delta\omega$ is actually the surface tension of the solid-vapor interface, $\gamma_{\sv}$, expressed explicitly as a function of its equilibrium film thickness, $h(t)$. By virtue of Eq.(\ref{eq:lndivg}), it follows that the Young condition for complete wetting is met exactly at the triple point. This serves to  emphasize the relation of complete surface melting with the wetting transition. But importantly, it shows explicitly that at temperatures below the triple point, a solid-vapor interface can exhibit a significant amount of interfacial premelting without it meeting exactly the wetting condition. This result is often not well appreciated.

Using the above results it is also possible to discuss what precisely can be
understood as the onset of premelting.\cite{qiu18,slater19,nagata19} The model
shows that premelting appears in a {\em continuous} fashion, as a gradual
flattening of the interfacial profile at the interface position $z=0$, i.e.,
there is no particular thermodynamic signature in the density profile of the
solid-vapor interface signaling a premelting threshold. The best that can be
done is to notice that the flattening of the interfacial density is the result
of the stabilization of the liquid phase at a metastable minimum of the free
energy surface. This is the only feature that  can be identified as the
theoretical onset of surface premelting. However,  the signature of the
metastable minimum is hardly detectable in the order parameter profile until the
minima of $f(\phi)$ become sufficiently close in energy, i.e. $t\approx 0$. Therefore, for practical matters  
the onset cannot be identified until the change in slope and the emerging liquid film at $z=0$ is above 
\change{the chosen experimental probe's detection threshold.}

\section{Wetting physics}

From this exactly solvable model, the continuous build up of a liquid film
between the solid and vapor phases, and its divergence at the triple point
appear as a very general feature of systems approaching three phase
coexistence. It is therefore tempting to consider complete surface melting as a
universal feature of solid-vapor interfaces.\cite{limmer14} However, the model
free energy of Eq.(\ref{eq:sft}) has  important limitations.

Firstly, it is based on a small-gradient, long wave-length approximation to the
free energy. This misses fine molecular details, such as packing correlations
between the repulsive core of the molecules,\cite{tarazona85} which can only be accounted with a multicomponent order parameter.\cite{lowen90,wang24} Secondly, it assumes implicitly that the molecular interactions are of short range nature. The model is therefore not able to account for the  role of long range interactions with algebraic decay such as van der Waals forces.\cite{schick90} 


In order to afford a more general description of interfacial premelting,  it is
convenient to borrow tools of  wetting physics.\cite{dietrich88,schick90} Here,
one is concerned with the behavior of an equilibrium solid-vapor interface as
the vapor phase approaches liquid-vapor coexistence. For the premelting of a
single component system, the condition of solid-vapor coexistence may be
achieved by moving along the sublimation line, while the approach to
liquid-vapor coexistence may be achieved by heating towards the triple point
along this line. Therefore, the framework of equilibrium wetting physics can be fully mapped
to the premelting problem {\em as long} as one moves along the sublimation line (Figure \ref{fig:phase_diagram} ).

 Let us then assume  a large parcel of ice is in phase equilibrium with an infinite amount of water vapor at temperature $T$ and vapor pressure $p_{\sv}(T)$.   This implies that, in a fixed sub-volume enclosing the ice-vapor  interface, the system is in a grand canonical ensemble, with a chemical potential that is set by the pressure of the water vapor. 
 This observation is convenient, because  the chemical potential is a scalar that remains constant throughout the system, and allows to characterize completely the thermodynamic state.\cite{evans79,henderson92} 
 As a bonus, the grand canonical free energy of a bulk system is merely given by $\Omega = -pV$, and is therefore the only free energy that can be  assessed from knowledge of mechanical properties solely. 
This allows us to estimate the free energy cost of replacing the bulk vapor with
a liquid layer off coexistence as $- \Delta p A h$, where $\Delta
p(T,\mu)=p_{\liq}(T,\mu)-p_v(T,\mu)$ is the pressure difference between bulk
liquid and vapor pressures at the temperature and chemical potential imposed by
the water vapor reservoir, and $A$ is the area of the interface.

The surface free energy of the solid-vapor interface per unit surface as a function of the film thickness can then be written as:\cite{llombart20,sibley21,baran24b}
\begin{equation}\label{eq:sfeh}
\Delta \omega(h) = \gamma_{\sl} + \gamma_{\lv} + g(h) - \Delta p h
\end{equation}
where $g(h)$ is  the {\em interface potential},  which measures the excess
surface free energy of the liquid film with respect to its bulk value; while $-\Delta p h$ accounts for the free energy cost of replacing a vapor layer  of thickness $h$  by equal amounts of  metastable bulk liquid. 

The fact that the surface free energy of the premelting film is measured with
respect to that of a bulk liquid does not imply  that the premelting film is
assumed to behave exactly as a liquid. Instead, by definition $g(h)$ is
introduced  here as a free energy excess with respect to the bulk liquid so as to account for such differences.  In practice, it is found that $g(h)$ becomes extremely small for films of barely one nanometer thickness.\cite{luengo22b} This shows that the bulk free energy of the liquid phase is indeed a good reference state for measuring the free energy of the film, and justifies the name of '{\em quasi-liquid layer}' that is often employed in place of 'premelting'.

Asides the quantitative measure of free energies, the role of $g(h)$ is to
describe the propensity of the premelting film to grow as three phase
coexistence is approached. By definition, $g(h)\to 0$ as $h\to \infty$.
Therefore, a monotonously decaying trend of $g(h)$ implies that the state of
minimal surface free energy corresponds to a bulk liquid film formed on the ice
surface, such that  the minimum  $h\to\infty$ . On the other hand, when $g(h)$ exhibits an absolute minimum at finite thickness, it corresponds to a case where the surface free energy is smaller for $h$ finite than $h\to\infty$, i.e. such that  the extent of premelting remains limited and there is therefore absence of complete surface melting. 

By comparing Eq.(\ref{eq:sfeh}) with Eq.(\ref{eq:sfew}), we see that  for the exactly solvable model of the previous section,  $g(h)=B e^{-\kappa_t h}$ is a monotonously decaying function corresponding to a case of complete surface melting.

Away from the triple point, with $\Delta p h$ finite, the interface potential
determines the equilibrium film thickness via the minimization of
Eq.(\ref{eq:sfeh}) with respect to $h$. This leads right away to:
\begin{equation}\label{eq:derjaguin}
\Pi(h_e) = -\Delta p
\end{equation}
where $\Pi=-dg(h)/dh$ is the {\em disjoining pressure}  as introduced by the
Russian school of colloidal science.\cite{derjaguin87,churaev88,henderson05} In
the geophysics community, $\Pi(h)$ is often known as the thermo-molecular force.\cite{wettlaufer99b,dash06}

This result shows that the film thickness may be considered as depending on the single variable  $\Delta p$, which characterizes the approach of the vapor to liquid-vapor coexistence. Since the vapor at pressures close to the sublimation line  behaves virtually as an ideal gas, $\Delta p(T)$ can be accurately approximated to  $\Delta p(T) = \rho_{\liq} k_BT  \ln p_{\sv}(T)/p_{\lv}(T)$,  
where $\rho_{\liq}$ is the density of the bulk liquid, $k_B$ is Boltzmann's constant and $p_{\alpha\vap}$ is the coexistence vapor pressure over $\alpha={\sol,\liq}$.\cite{llombart20,sibley21}
This  shows explicitly  that the equilibrium premelting thickness need not be measured strictly in terms of a distance away from the triple point. Rather, it is dictated by the distance of the sublimation line to the metastable prolongation of the condensation line. In water, these two lines run very close together all the way down to 0~K, an observation that could explain the unexpected report of ice premelting at very low temperature.\cite{lin23}   Assuming Clausius-Clapeyron behavior for the vapor pressures, this equation simplifies to $\Delta p=\rho_l \Delta h_{\sl} (T-T_t)/T_t$, where $\Delta h_{\sl}$ is the enthalpy of melting.  The result is quantitatively correct down to $10^{\circ}$~C, but remains qualitatively correct for at least $-50^{\circ}$~C.\cite{baran24b}

Unfortunately, the interface potential is usually not known exactly. Under this circumstances, Eq.(\ref{eq:derjaguin}) is not a predictive tool, but rather, a practical means for obtaining information of the related disjoining pressure from knowledge of the equilibrium film thickness.\cite{llombart20,sibley21,baran24b,baran25} 
Thus, if one has reliable measurements of $h_e$ as a function  of temperature, together with computations of $\Delta p$ from bulk thermodynamics, Eq.(\ref{eq:derjaguin}) provides the disjoining pressure right away.\cite{llombart20,sibley21,baran24b}

The reader might wonder what is the point of mapping $h_e$ to $\Pi(h)$ via $\Delta p$, if the experimental measurements provide right away $h_e$ as a function of temperature, which is a simpler variable. The answer is that, once $\Pi(h)$ is known,  Eq.(\ref{eq:derjaguin})
shows that $h_e$ is a natural function of the single variable $\Delta p(T,p)$.
This allows to predict equilibrium thickness for arbitrary thermodynamic states
from knowledge of a set of measured states. For the case of surface premelting,
knowledge of $\Pi(h)$ obtained for equilibrium states along the sublimation
line, allows to predict the properties of premelting films in moderately out of 
equilibrium conditions of arbitrary temperature and pressure, as we shall discuss 
later in Section \ref{sec:noneq}.\cite{sibley21} 

\begin{figure}
\includegraphics[width=0.55\textwidth]{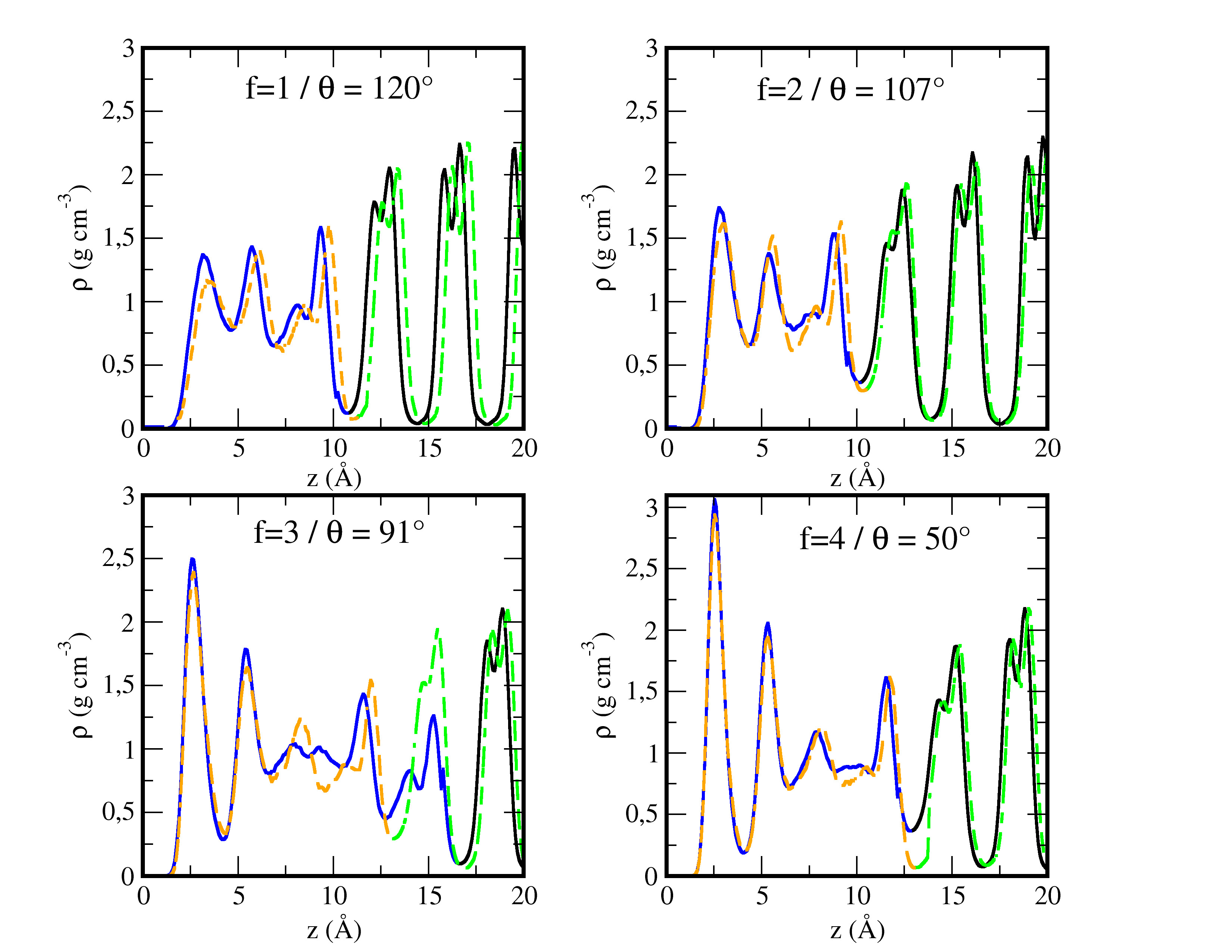}
\caption{Characterization of film structure on substrates of different
wettability as indicated by the contact angle, $\theta$. Interfacially premelted
films in widely different conditions but equal disjoining pressure are very
similar. The density profiles shown in dashed lines  were obtained at $T=266$~K and $p=1$~bar; The ones in a continuum line are obtained at $T=262$~K and $p=400$~bar.  The region of density profiles with a majority of liquid like molecules is depicted in either orange or blue; that where the majority is formed by solid like molecules is shown in green and black. Reproduced with permission from Baran et al.\cite{baran25} J. Phys. Chem.~C {\bf 129} 4614 (2025). Copyright \copyright  2025 American Chemical Society. \label{fig:mapping}}
\end{figure}

This concept may be easily tested without invoking out of equilibrium conditions for the closely related case of interfacial premelting, where premelted equilibrium states can now occur for arbitrary values of temperature and pressure.\cite{baran25}  In this context, Eq.(\ref{eq:derjaguin}) allows to predict equilibrium film thicknesses at arbitrary thermodynamics states from measurements of film thickness along a single isobar. This shows that the natural measure of distance away from a point in the melting line $(T_m,p_m)$ is neither $T-T_m$, nor $p-p_m$, but the pressure difference $\Delta p(T,p)$.
In a recent study, it was shown that this mapping  holds also for  the prediction of the interfacial structure of the premelting film, as illustrated in Fig.\ref{fig:mapping}.\cite{baran25}

\section{Surface Intermolecular Forces}

The other important reason why the wetting physics formalism can proof as a formidable benchmark for understanding ice premelting is that the interface potential $g(h)$  has its origin in molecular interactions. Therefore, one can use a large body of well understood results on the theory of intermolecular forces,\cite{israelachvili11,butt10} to constraint $g(h)$ and make important qualitative statements about its behavior.\cite{luengo22b}

The first simple insight one can make is that the interface potential of a premelting film may be divided into short and long range contributions additively as:
\begin{equation}\label{eq:gtot}
g(h) = g_{sr}(h) + g_{vdw}(h)
\end{equation}
Here, $g_{sr}(h)$ is a short range contribution with exponentially fast decay which results from packing correlations,\cite{chernov88,henderson94,leote94,henderson05} short range attractive interactions,\cite{derjaguin87,israelachvili11} and finer effects such as renormalizable surface fluctuations.\cite{chernov88,henderson05} Additionally, $g_{vdw}(h)$ is a long range contribution that stems from the van der Waals interactions and decays in algebraic fashion.\cite{parsegian70,parsegian05,israelachvili11,dietrich91}

\subsection{Packing and layering effects}

\begin{figure*}[t]
	\centering
	\begin{minipage}[b]{0.50\textwidth}
		\begin{overpic}[width=\textwidth]{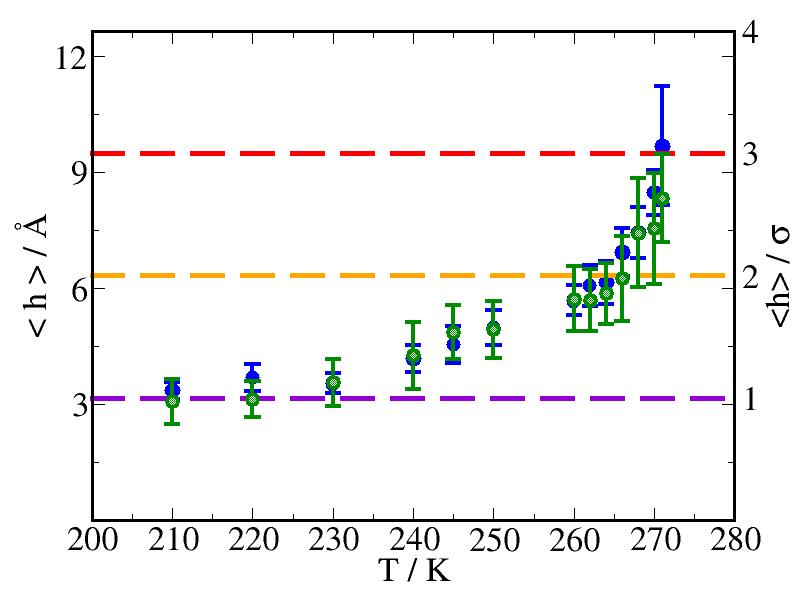}
			\put(15,80){\textbf{(a)}} 
		\end{overpic}
	\end{minipage}%
	\hfill
	\begin{minipage}[b]{0.50\textwidth}
		\begin{overpic}[width=\textwidth,height=0.36\paperwidth]{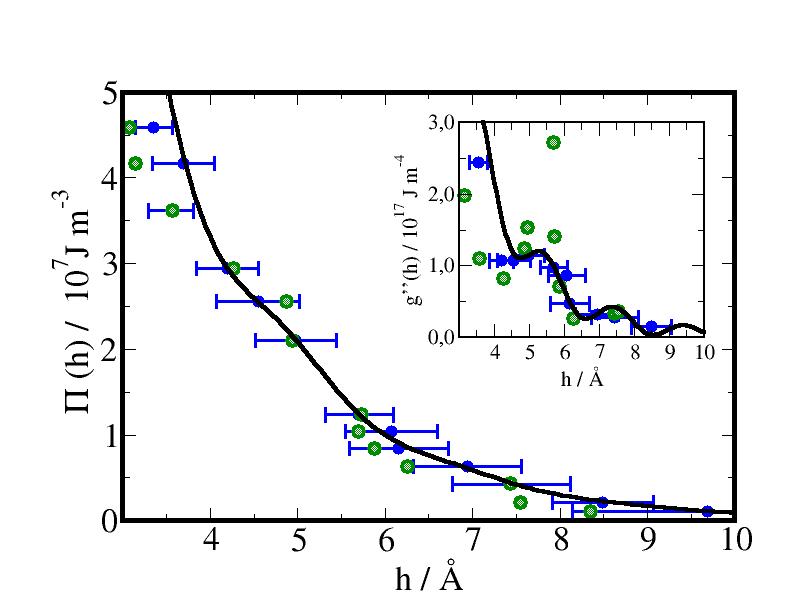}
			\put(15,80){\textbf{(b)}} 
		\end{overpic}
	\end{minipage}
	\caption{Calculation of disjoining pressures. a) Film thickness measured at
		solid-vapor equilibrium for the primary prism face as a function of temperature. b) The
		results are mapped to the corresponding Laplace pressure difference,
		$\Delta p$,
		between bulk water and vapor at the same chemical potential to produce a
		disjoining pressure isotherm. Symbols are results from computer
		simulations for the TIP4P/Ice model. The lines are a fit to the
		asymptotic result of Eq.(\ref{eq:gsr}). The inset shows inverse
		susceptibilities as calculated from the simulations (symbols) and
		predicted from the second derivative of the interface potential
		(lines), displaying regions of enhanced surface fluctuations. Reproduced with permission from Llombart et al.\cite{llombart20} Phys. Rev. Lett. {\bf 124} 065702 (2020). Copyright \copyright 2020 American Physical Society.    	
		\label{fig:disjoining}}
\end{figure*}

Borrowing tools from liquid state theory, the short range contribution is expected to obey the following equation:\cite{chernov88,henderson94,leote94,henderson05}
\begin{equation}\label{eq:gsr}
g_{sr}(h) = B_2 e^{-\kappa_2 h} - B_1 e^{\kappa_1 h} cos(q_o h - \alpha)
\end{equation}
where the $B_i$ are positive coefficients, $\kappa_i$ are inverse decay lengths in the scale of the molecular diameter, $q_o$ is a wave-vector which sets the pitch of the packing correlations, and $\alpha$ is an unknown phase. 

In this expression, the first term is a monotonous decaying function which
reduces the film's free energy as $h$ increases. Accordingly, this term promotes
surface melting, as predicted by the toy model of Section III and  previous studies.\cite{chernov88,limmer14} 
To a higher order of approximation, however, it is required to account for
the packing correlations of the water molecules, which produce
next to leading order contributions of damped oscillatory form.\cite{tarazona85}
At the mean field level, the presence of this term suggests that the interface potential could exhibit
local minima, and generate a sequence of layering
transitions.\cite{chernov88} At a higher level of approximation,
surface capillary waves can renormalize the mean field expectation and suppress
the predicted layering transitions.\cite{chernov88,henderson05}

Figure~\ref{fig:disjoining} shows an example for the disjoining pressure curve $\Pi=dg/dh$ 
\change{of the prism face} as obtained from computer simulations using the TIP4P/Ice model of
water.\cite{abascal05} The curve is computed by first measuring the equilibrium premelting thickness as a 
function of temperature (Fig.\ref{fig:disjoining}-a),  mapping this to the corresponding value of $\Delta p(T)$, 
as obtained accurately from thermodynamic integration, and plotting $\Delta
p(T)$ as a function of $h_e(T)$ (Fig.\ref{fig:disjoining}-b).\cite{llombart20}

The results show a monotonous decrease of the disjoining pressure. Therefore,  the amplitude of the oscillations predicted in Eq.\ref{eq:gsr} is too small to allow for
the appearance of minima of the interface potential for films of less than
several nanometers. In this fluctuation
dominated regime, the oscillations predicted in the mean field approximation
are washed away, and there is no longer a `layering transition' in the
thermodynamic sense.\cite{chernov88,henderson05} However, the signature of
underlying mean field transitions can survive as a flattening of the interface
potential within given ranges of film thickness. In the computer simulations of
Ref.\cite{llombart20}, \change{these} rounded transitions could be observed as a marked
oscillatory behavior of the inverse interfacial susceptibility (inset of
Fig.\ref{fig:disjoining}-b), which can be measured as the second derivative of the interface potential. 

Interestingly, experimental studies using Sum Frequency Generation spectroscopy observed a stepwise blue shift in the OH stretch signal, which was interpreted as a hint of  layer by layer growth at the basal surface of ice at about 257~K.\cite{sanchez17,michaelides17} According to the computer simulations, this signal could correspond to enhanced surface fluctuations at a rounded layering transition, but the current simulation results suggest this is not a strict thermodynamic transition. Unfortunately, surface phase transitions in the fluctuation dominated regime exhibit strong system size effects, and further simulations with larger systems might be necessary to settle the issue.

To summarize, \change{all computer simulation studies support the emergence and gradual growth of premelting films, with a film thickness of about 1 to 2 nm one Kelvin away from the triple point.}\cite{conde08,limmer14,llombart19,llombart20,llombart20b,sibley21,baran24b}
A detailed analysis suggests that in this range, there are enhanced surface fluctuations on growing from one bilayer to the next, but not a thermodynamic layering transition of the premelting film in the thermodynamic sense. Extrapolation of the fits  beyond the range of film thicknesses achieved in simulation suggest that the  oscillatory term of Eq.(\ref{eq:gsr}) could in principle lead to very shallow minima of the interface potential for the basal face,\cite{llombart20,baran24b} but not for the prism face. However, the resulting minima are so shallow that they are very likely to vanish due to surface capillary waves. Under this assumption, the short range contribution of the interface potential is a monotonously decreasing function, which effectively promotes a situation of complete surface melting for both the basal and prism face, in a way similar to the prediction of the toy model of Eq.\ref{eq:sfew}.

\subsection{Van der Wals forces and the absence of complete melting}

The above discussion suggests that, as long as short range forces are concerned, ice should exhibit complete surface melting. However, as shown in Eq.(\ref{eq:gtot}), the interface potential also contains contributions that result from the van der Waals interactions between molecules. These forces have an algebraic decay, so that, for sufficiently thick premelting films they will dominate over the short range forces, which decay exponentially fast. Therefore, the ultimate fate of the interface potential for thick wetting films will be dictated, not by the short range forces, but by the weak van der Waals forces.

Unfortunately, it is not possible to assess directly the role of van der Waals
forces by means of computer simulations for two reasons. The first and most
important is that the van der Waals forces in computer simulations are cutoff
beyond a truncation radius of about one nanometer, and this has been no
exception in the study of ice premelting. The second is that, even if the
interactions were calculated effectively, say, as in a lattice
sum,\cite{karasawa89} the accuracy required to measure the effect of such forces would be likely prohibitive. Among other things, reaching a thick film beyond the nanometer requires approaching very carefully the triple point in the sub-Kelvin scale, but in practice, the uncertainty in the triple point of usual force fields is in the scale of one to two Kelvin.

Fortunately, the nature of surface van der Waals forces is rather well understood and allows to constrain the interface potential without the need of involved computer simulations. Indeed, borrowing the tools of surface physics, we can describe the role of van der Waals forces on a wetting film by means of the following equation:\cite{dzyaloshinskii61,parsegian70,parsegian05}
\begin{equation}\label{eq:Adeh}
   g_{vdw}(h) = -\frac{A(h)}{12\pi h^2}
\end{equation}
where $A(h)$ is the so called Hamaker constant. The full account of the $h$
dependent Hamaker constant requires very involved quantum-electrodynamic
considerations,\cite{parsegian05} but to the accuracy that can be achieved in
current experiments, we can simplify the problem to the range of $h$ less than
100~nm, where $A(h)$ depends only weakly on the film thickness.\cite{luengo22b}

In this situation, corresponding to $A(h\to0)=A$, the van der Waals forces are fully dominated by high frequency dipole fluctuations of the electronic degrees of freedom. These, in practice, are modeled in computer simulations and most theoretical approaches by a dispersive London term between water molecules at the pair level, which decays as $-C_{vdw}/r^{6}$. Such contributions can be summed up in the pair additive approximation of Hamaker,\cite{hamaker37} that has been revisited more recently under the framework of (thermodynamic) density functional theory of wetting films.\cite{schick90,dietrich91,macdowell17} For the special case of surface premelting, the result reduces to:\cite{benet19,luengo22b,baran24b}
\begin{equation}\label{eq:ska}
  A = \pi^2 C_{vdw}(\rho_{\vap}-\rho_{\liq})(\rho_{\sol}-\rho_{\liq})
\end{equation} 
For most substances, the density of the solid phase is larger than that of the
liquid phase, and the Hamaker constant is negative. In such case, the van der
Waals contribution to the interface potential becomes positive, and favors complete surface melting. However, a well known anomaly of water is that   the density of ice is smaller than that of liquid water. This means that the Hamaker constant 
ruling ice premelting is positive, a fine detail that was already noticed by
Nozieres.\cite{nozieres92} The implication is  that the van der Waals forces take a negative value, and inhibit the formation of a thick wetting film.  So, whatever is the behavior of the short range contribution, and whether it adopts minima due to the oscillations of the packing term or not, the contribution of the van der Waals forces is to inhibit the growth of the premelting film. As a result of this balance, the interface potential must have a minimum at intermediate distances, leading to a case of incomplete surface melting. \change{Further notice that this  behavior is generic and arises from the nature of the bulk ice and water phases, so that, to leading order, it does not depend on the nature of the exposed surface plane}. 

\subsection{A minimal model of ice premelting}

\begin{figure}
	\includegraphics[width=0.5\textwidth]{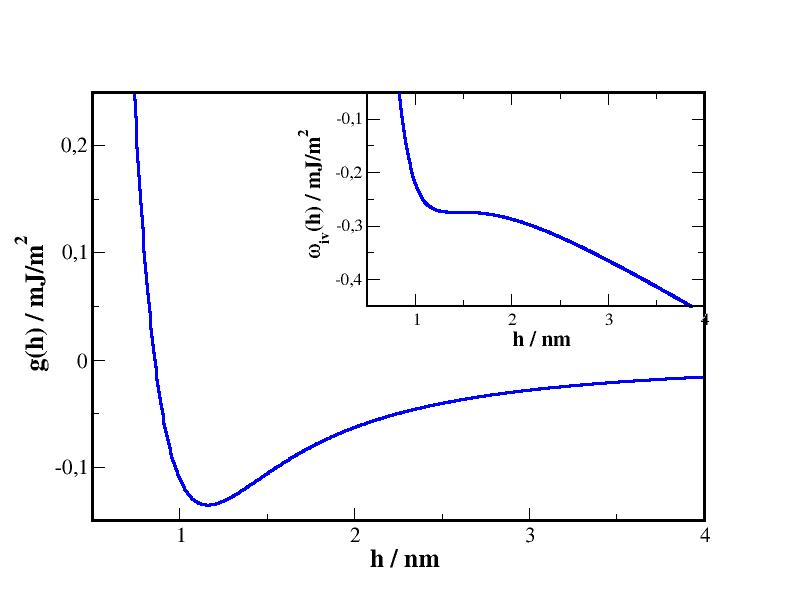}
	\caption{A minimal model of the interface potential of ice premelting as
obtained from a single decaying exponential function and a constant positive
Hamaker constant. The minimum of the interface potential leads to an equilibrium
premelting film thickness at the triple point of about 1~nm. Inset: Away from
the triple point, the surface free energy of Eq.(\ref{eq:sfeh}) picks up the $-\Delta p h$ term,
until the minimum vanishes at a wetting spinodal point.
	Reproduced from  Luengo et al.\cite{luengo22b} J. Chem. Phys. {\bf 157} 044704 (2022), with the permission of AIP Publishing.
		\label{fig:gminmod}}
\end{figure}

Based on the discussion in Sections V.B and V.C, a minimal model of ice premelting needs to account for the growth of the premelting film due to short range structural forces, and the inhibition of further growth by long range van der Waals forces. Ignoring the small oscillatory term, such behavior can be described  with the following  interface potential:
\begin{equation}
   g(h) = B e^{-\kappa h} - \frac{A}{12\pi h^2}
\end{equation}
The result obtained by parametrization to data for the primary prism face (Figure~\ref{fig:gminmod}) shows a shallow minimum at barely one nanometer thickness, which inhibits complete surface melting at conditions of solid-vapor coexistence.\cite{luengo22b} However, as soon as the system is moved away from the sublimation line, $\Delta p$ in Eq.(\ref{eq:gtot}\change{)}  can increase sufficiently that the overall free energy minimum vanishes completely at a {\em wetting spinodal point} (Figure~\ref{fig:gminmod}-inset). Eventually, this can lead to the complete wetting of the system as discussed in Section \ref{sec:noneq}.

\subsection{Disgression: Into the realm of quantum electrodynamics}

\begin{figure}
	\includegraphics[width=0.5\textwidth,keepaspectratio]{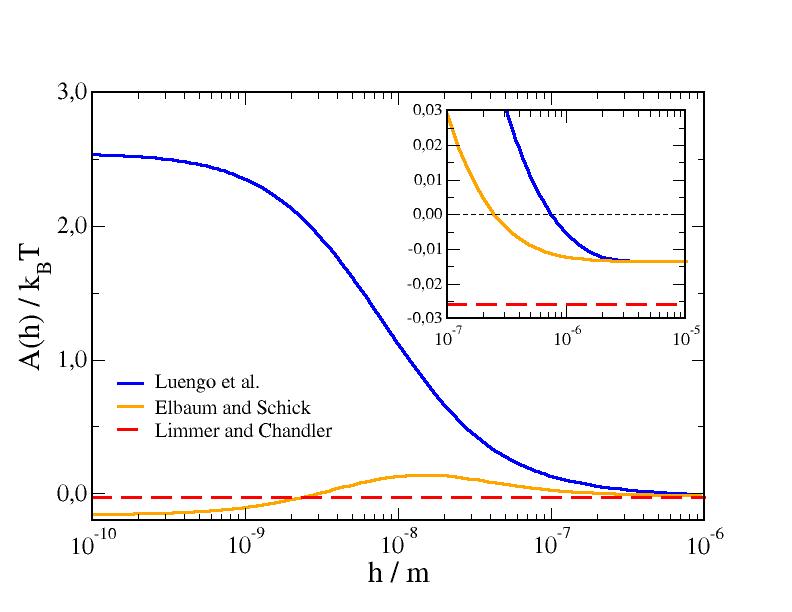}
	\caption{{ Hamaker functions as predicted by DLP theory}. Blue lines are results for $A(h)$ as calculated from an accurate parametrization of the dielectric response of water and ice by Luengo et al.\cite{luengo22b}. For small $h$, $A(h)$ is positive as predicted by Eq.(\ref{eq:ska}), and inhibits surface melting. For large $h$, $A(h)$ changes sign, becomes negative and favor surface melting. The inset shows the sign reversal occurs at about 760~nm. \change{Orange} lines are results from Elbaum and Schick, which suffer from a poor parametrization of the dielectric response.\cite{elbaum91b} The sign of $A(h)$ is correct for films of thickness larger than 3~nm. The red dashed lines are results from Limmer and Chandler.\cite{limmer14} 
	Adapted  from  Luengo et al. J. Chem. Phys.\cite{luengo22b} {\bf 157} 044704 (2022), with the permission of AIP Publishing.
		 \label{fig:hamaker}
		 }
\end{figure}

The minimal model discussed above describes the qualitative behavior of ice premelting for film thicknesses less than hundreds of nanometers. Beyond this range, intermolecular forces become dominated by very subtle quantum electrodynamic effects, that are usually ignored altogether.\cite{israelachvili11} Indeed, as emphasized in Eq.(\ref{eq:Adeh}), the Hamaker 'constant' is actually not a constant, but displays a very rich and complex behavior as the thickness of the premelting film becomes large. The full quantitative behavior of $A(h)$ can only be determined within the framework of  Dzyaloshinskii-Lifshitz-Pitaevskii theory (DLP),\cite{dzyaloshinskii61} which has been used extensively by Bostr\change{\"o}m and collaborators to address a number of related problems in ice premelting.\cite{bostrom17,bostrom19,esteso20,luengo21,yang22,bostrom23} Unfortunately, DLP theory is not amenable to an easy qualitative interpretation, and requires involved numerical calculations.\cite{parsegian05} For this reason, a simple analytical approximation  was worked out recently solely for this purpose.\cite{macdowell19}  

The rationale here is that van der Waals forces emerge from dipole fluctuations
at all frequencies of the spectral range. The Hamaker function $A(h)$ is
therefore a sum of contributions of all frequencies. In practice, for films
smaller than a hundred nanometers or so, the Hamaker function is fully
dominated by high frequency dipole fluctuations, and is approximately given
as:\cite{macdowell19}
\begin{equation}\label{eq:awfh0}
   A_{\omega > 0}(h)   =  \displaystyle{ \frac{3\hbar c \nu_{\infty}}{32\sqrt{2}\,n_{\liq} }
   	\left ( \frac{n_{\sol}^2-n_{\liq}^2}{n_{\sol}^2+n_{\liq}^2} \frac{n_{\vap}^2-n_{\liq}^2}{n_{\vap}^2+n_{\liq}^2}
   	\right ) } f(h)
\end{equation}
where $\hbar$ is Planck's constant, $c$ is the speed of light,
$\nu_{\infty}\approx 0.01$~nm$^{-1}$ is a wave-number corresponding to the principal
electronic transition of water, $n_i$ are refractive \change{indexes}  in the
visible, and finally, $f(h)$ is a crossover function which remains close to
unity for all $h\nu_{\infty}\ll 1$. In this regime of small $h$, the above equation
accounts for correlated electronic dipole fluctuations as dictated by the
material's refractive indexes. 
As an anomaly of water, the refractive index of the solid phase is smaller than
that of the liquid phase, $n_{\sol}<n_{\liq}$, which results in   a positive
Hamaker constant, exactly as predicted by Eq.(\ref{eq:ska}).

However, as the film thickness gradually increases,  quantum-electrodynamic
effects gradually suppress the contributions from high frequency dipole
fluctuations, leading to the vanishing of the crossover function approximately
as:\cite{macdowell19}
\begin{equation}\label{eq:awf}
f(h)=  \frac{1}{\nu_{\infty}h}
\left [
(2+\frac{3}{2}\nu_T h) e^{-\nu_T h} - (2 + \nu_{\infty} h ) e^{-\nu_{\infty}
	h}
\right ]
\end{equation}
where $\nu_T$ is a thermal wave-number in the micrometer scale. From this
equation it is possible to assess easily that for $h\nu_{\infty} \ll 1$,
$f(h)\approx 1$, which justifies the Hamaker approximation for sufficiently
thin films. When the film thickness increases such that $h \nu_T < 1 < h\nu_{\infty}$, 
$f(h)$  gradually decreases as $1/h$, leading to the regime of retarded Casimir
interactions. Finally, when $h\nu_T>1$, the crossover function vanishes
exponentially fast, leading to the complete suppression of high frequency
contributions to the intermolecular forces. In this little appreciated regime,
the Hamaker function now becomes dictated by static contributions alone, and is
given by:
\begin{equation}\label{eq:aw0}
     A_{\omega=0}= \frac{3}{4}  
     \frac{(\epsilon_{\sol} -\epsilon_{\liq})}{(\epsilon_{\sol}+\epsilon_{\liq})}
     \frac{(\epsilon_{\vap} -\epsilon_{\liq})}{(\epsilon_{\vap}+\epsilon_{\liq})}
     k_B T 
\end{equation}
where, relative to Eq.(\ref{eq:awfh0}), the dipole fluctuations are now described in
terms of the static dielectric constants, $\epsilon_i$. Since in this case
$(\epsilon_{\liq}-\epsilon_{\sol})<0$, the change from high frequency dominated to low frequency dominated 
behavior leads to a sign reversal of the Hamaker function, which  changes from being positive for $h$ small, 
as anticipated by  Eq.(\ref{eq:ska}), to being negative for $h$ larger than the micrometer scale. 
This means that $A(h)$ shifts from inhibiting surface melting at $h<\mu$m,  to  promoting surface melting at 
large $h>\mu$m.\cite{luengo22b,fiedler20}

In practice, whether one considers the simple description of Eq.(\ref{eq:ska})
or the complex $A(h)$ function from DLP theory, the interface potential exhibits
an absolute free energy minimum at intermediate film thicknesses, so that the equilibrium
state of the ice surface at the triple point corresponds to a case of incomplete
surface melting. If, however, for whatever reason a large fluctuation moves the
film thickness into the range where $A(h)$ has become negative, the  surface
intermolecular forces could lead the system to surface melt spontaneously.

Elbaum and Schick pioneered the use of DLP theory as a tool for the study of ice premelting.\cite{elbaum91b} Their results anticipated that van der Waals forces inhibit ice premelting for films in the nanometer scale.  Unfortunately, they used a poor parametrization of the dielectric functions of water and ice, and predicted erroneously a negative Hamaker constant promoting surface premelting as $h\to 0$.  To add to this confusion, Limmer and Chandler approximated the Hamaker constant at small $h$ using the static contribution that is relevant only at large $h$, but predicted the correct behavior due to an error in the sign convention in Eq.(\ref{eq:Adeh}).\cite{limmer14} 

The problem has since then been revisited, and new, more reliable parametrizations of the dielectric response confirm that the Hamaker function is positive in all the range from $h=0$ to the micrometer scale as described above.\cite{luengo22b,fiedler20} Unfortunately, most of the work on DLP performed previous to 2020,  including also work by myself,\cite{esteso20} has relied on the parametrization of Elbaum and Schick, and could be unreliable for premelting films smaller than the nanometer. 

Figure \ref{fig:hamaker} shows the full $h$ dependence of the Hamaker function
with a recent reliable parametrization of the dielectric
properties,\cite{luengo22b} and compares it to previous
calculations.\cite{elbaum91b,limmer14} The results of Luengo et al. are in
qualitative agreement with  Fiedler et
al.\cite{fiedler20} (not shown), but  Bostr\change{\"o}m et al. have ever since opted for the parametrization 
of dielectric properties presented in Ref.\cite{luengo22b}.

\section{Water droplets on ice}

\label{sec:drops}

\begin{figure*}[t]
	\centering
	\begin{minipage}[b]{0.32\textwidth}
		\begin{overpic}[width=\textwidth]{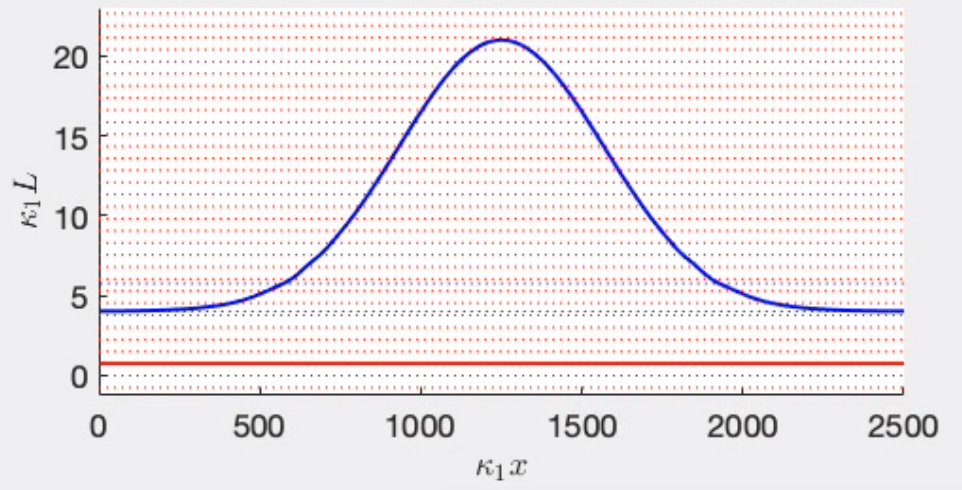}
			\put(15,44){\textbf{(a)}} 
		\end{overpic}
	\end{minipage}%
	\hfill
	\begin{minipage}[b]{0.32\textwidth}
		\begin{overpic}[width=\textwidth]{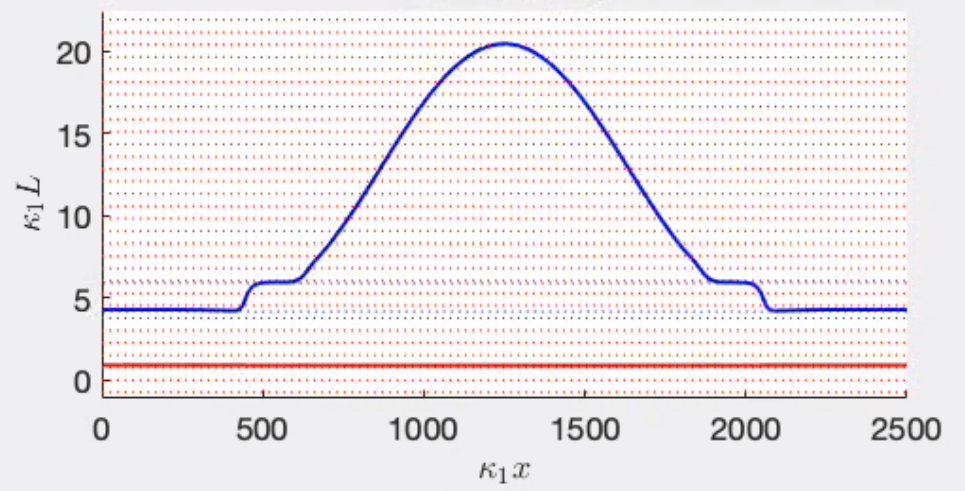}
			\put(15,44){\textbf{(b)}} 
		\end{overpic}
	\end{minipage}%
	\hfill
	\begin{minipage}[b]{0.32\textwidth}
		\begin{overpic}[width=\textwidth]{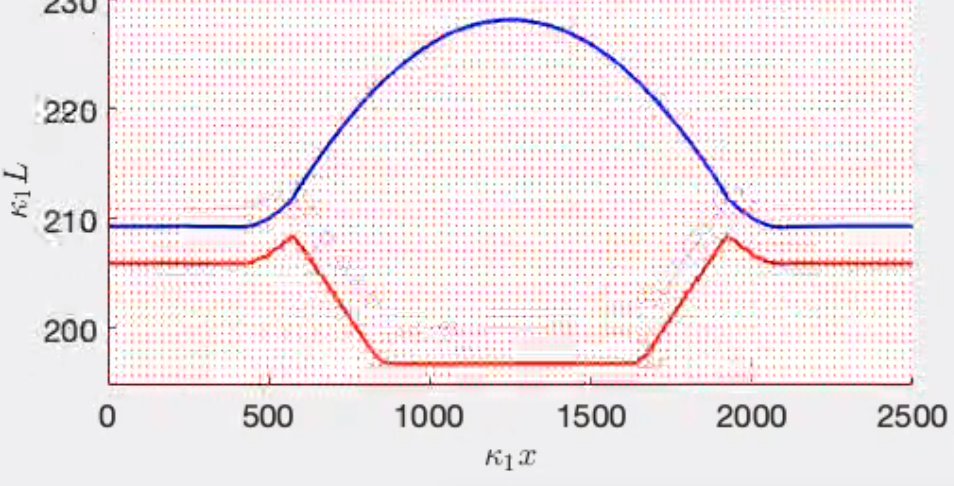}
			\put(15,44){\textbf{(c)}} 
		\end{overpic}
	\end{minipage}
\caption{Mesoscopic modeling of water droplets on ice. The figure shows the time evolution of a liquid droplet deposited at conditions of supersaturation over water (kinetic condensation line). The blue line depicts the liquid/vapor surface, while the red line describes the position of the ice/water surface. a) Notice the drop sits on top of a premelting film. b) Transiently, droplets governed by an interface potential with two minima can exhibit a second film thickness next to the rim of the drop. c) For sufficiently long times, as water vapor condenses the surfaces elevate by condensation and ice gradually freezes within the drop. Reproduced with permission  from Sibley et al.\cite{sibley21} Nat. Comm.  {\bf 12} 239 (2021).  Distributed under a Creative Commons Attribution License CC-BY \label{fig:mesodrop}}

\end{figure*}

As discussed above, the theoretical considerations show that pure ice does not
exhibit complete surface melting. However, the experimental measurements of
premelting thickness are far less clear  and often appear to favor the opposite
conclusion, as evidenced from the positive slopes of $h(T)$ in a log-log plot
(Fig.\ref{fig:hvT}). Yet, for a long time there has been compelling experimental  evidence that water does not completely wet the ice surface. This evidence comes from the recurrent observation of water droplets formed on its surface, first reported by Knight.\cite{knight67,knight71} 

The presence of water droplets has ever since been confirmed by a number of
studies through the years,\cite{ketcham69,elbaum91,elbaum93} including
photographs of water droplets on the basal surface of hexagonal ice
micro-crystals,\cite{gonda99} and large millimeter drops on polycrystalline
ice.\cite{makkonen97,thievenaz20,demmenie25,sarlin25}  However, it was not until \change{the last}
ten years that the quality of the reported pictures has allowed to appreciate this finding widely.\cite{sazaki12,asakawa16,murata16,sazaki22}

According to Young's equation, the surface tension of a solid-vapor interface is
related to the equilibrium contact angle as $\gamma_{\sv} = \gamma_{\sl} +
\gamma_{lv}\cos\theta_e$. This means that droplets with a finite contact angle,
$\theta_e>0$, have a surface tension less than $\gamma_{\sl}+\gamma_{lv}$, and
do not meet Young's condition of complete wetting. Therefore, the observation of water droplets implies 
necessarily a case of incomplete surface melting. 

How can then we reconcile the experimental observations of incomplete wetting with the reported divergence of premelting film thickness as temperature approaches the triple point? 

The key point as suggested by Sazaki and collaborators,\cite{sazaki12,asakawa16} is that conventional techniques such as x-ray dispersion,\cite{dosch95} spectroscopy,\cite{bluhm02,sadtchenko02} or ellipsometry,\cite{furukawa87} do not allow to tell whether the water like signal of the experimental probe is due to the continuous growth of a uniform liquid layer, or to the spatially heterogeneous spreading of liquid droplets. In fact, studies by Elbaum et al.\cite{elbaum93} are likely the only ones to have combined simultaneously a reflectometer to measure film thickness and an interference microscope to visualize the ice surface. In their experiments, they could confirm that close to the triple point, the gradual growth of the premelting film thickness was eventually followed by the formation of condensed water droplets which produced an apparent diverging film thickness.\cite{elbaum93} This shows that the use of conventional techniques becomes unreliable in the close neighborhood of the triple point.The reason is that the condensation and sublimation lines are very close to each other and eventually meet exactly (Figure \ref{fig:phase_diagram}). Therefore, it suffices a minimal local fluctuation for the vapor to become supersaturated over water and move the solid-vapor interface away from equilibrium conditions. This results in vapor condensation, and the appearance of inhomogeneities and instabilities. In such conditions, the ice-vapor interface does no longer correspond to an equilibrium uniform state, and the measured water-like signal can no longer be trusted as a measure of the equilibrium film thickness.

The take away message from the experimental side is that, 1) experimental probes of film thickness confirm the formation of liquid-like layers on the ice surface and 2) contact angle measurements confirm that ice exhibits only incomplete surface melting. This scenario is consistent with observations from computer simulations and agrees with the picture that emerges from the physics of intermolecular forces,\cite{nozieres92,elbaum91b,benet19,fiedler20,sibley21,luengo22b} although a full understanding of the origin of the incomplete surface melting has emerged only recently.\cite{llombart20,luengo22b}

Paradoxically, the experimental observation of liquid droplets on the ice surface has often been interpreted as illustrating the complete lack of equilibrium ice premelting,\cite{knight96,makkonen97,sazaki12,asakawa16,murata16,sazaki22,demmenie25}
which is obviously at odds with experimental measurements of film
thickness.\cite{elbaum93,dosch95,sadtchenko02,mitsui19} This conflict emerges
from a misunderstanding of Young's equation, where it is often assumed that
$\gamma_{\sv}$ accounts for the surface tension of a {\em bare} interface
between the unrelaxed ice surface and its vapor. 
At high temperature, the {\em equilibrium}
solid-vapor interface can exhibit a significant amount of surface disorder 
consistent with the existence of a premelting film, and $\gamma_{\sv}$ may
become significantly different from the expected value for a perfect unrelaxed
interface obtained by cleavage of the solid phase.

The key issue here is to realize that the surface free energy of Eq.(\ref{eq:sfeh}) becomes the exact solid-vapor surface tension when evaluated for the equilibrium film thickness, $h_e$, as dictated by Eq.(\ref{eq:derjaguin}).  Indeed, replacing the equilibrium condition into Eq.(\ref{eq:sfeh}) must provide the ice-vapor surface tension by definition, whence:
\begin{equation}\label{eq:svifp}
\gamma_{sv} = \gamma_{\sl} + \gamma_{\lv} + g(h_e) + \Pi(h_e) h_e
\end{equation} 
Akin to the result of the exactly solvable model, Eq.(\ref{eq:sfew}), this equation shows explicitly that the ice-vapor interface in thermodynamic equilibrium can exhibit an adsorbed liquid film of finite thickness without necessarily meeting the Young condition for wetting. Particularly, equating Eq.(\ref{eq:svifp}) to Young's condition for a droplet of liquid on a solid substrate, one gets an equation for the contact angle  at the triple point:
\begin{equation}\label{eq:cag}
\cos(\theta_e) = 1 + \frac{g(h_e)}{\gamma_{lv}} 
\end{equation}
This explicitly provides the contact angle of a droplet formed atop a premelting
film of equilibrium thickness $h_e$ (c.f. Figure \ref{fig:mesodrop} for the numerical calculation of a water droplet profile consistent with Eq.(\ref{eq:cag})).

This now shows again that observation of water droplets on the ice surface found
in many different experiments over the years,\cite{knight67,ketcham69,knight71,gonda99,elbaum91,elbaum93,gonda99,sazaki13,murata16,demmenie25,sarlin25} need not rule out the presence of an equilibrium wetting film of finite thickness below the liquid droplet as occasionally assumed.\cite{knight96,makkonen97,sazaki12,asakawa16,murata16,sazaki22,demmenie25}

The confusion is likely to result from an alternative \change{convention}  for the surface free energy which is often found in the literature:\cite{dash06,nagata19}
\begin{equation}\label{eq:theta}
\Delta\omega(h) = (\gamma_{\sl}+\gamma_{\lv}-\gamma_{sv}) I(h) + \gamma_{sv} - \Delta p h
\end{equation}
\change{Comparing this result with Eq.(\ref{eq:sfeh})}, one can easily relate $I(h)$ with the  interface potential, $g(h)$, whereupon, both \change{results} are fully equivalent. However, assuming that $I(h)=0$ for $h=0$ from the outset,\cite{dash06,nagata19} leads to the misleading interpretation that the solid-vapor interface need be understood as a bare ice-vapor interface with no premelting whatsoever. Actually, the above equation is completely equivalent to Eq.(\ref{eq:sfeh}), with the understanding that $I(h)=0$ holds not for $h=0$, but for $h=h_e$.

It is noteworthy that the calculations of interface potentials as described
above predict a shallow minimum at an equilibrium film thickness slightly above
$h_e\approx 1~nm$,  with $g(h_e)$ in the scale of
$-10^{-1}$~mJ/m$^2$.\cite{luengo22b,baran24b} Using \change{Eq.(\ref{eq:cag})}, this
shows that atop the premelting film of nanometer thickness, a water droplet with
contact angles ranging from 2.5 to 3.5 degrees can form. This appears to be in
rather fair agreement with observation of water droplets formed
spontaneously by condensation or melting on pristine ice facets at very nearly
equilibrium conditions, which are reported to fall in the range of 0.6 to 2.3
degrees.\cite{ketcham69,elbaum93,murata16,nagata19,sazaki22} 

Intriguingly, experiments  performed with macroscopic droplets  on the order of one millimeter radius provide significantly larger contact angles.\cite{knight67,makkonen97,demmenie25,sarlin25} In two recent experiments,\cite{demmenie25,sarlin25} liquid water droplets above the melting point were gently deposited on ice, and the relaxation process was followed carefully until full arrest of the contact line.  For ice at a temperature close to its melting point, these experiments  reported consistently contact angles of 12$^{\circ}$, in agreement with the early observation of Knight.\cite{knight67} However,
the deposition process here is an extremely complex problem in fluid mechanics,
as the droplets are formed  in conditions of significant temperature and
chemical potential  gradients between the outer atmosphere, the water droplet
and the underlying ice. So despite that the relaxation process was carefully
followed and checked for the attainment of {\em mechanical equilibrium}, it is
unlikely that this corresponds to the conditions of experiments where spontaneous
droplet formation was observed by gentle condensation of the vapor at nearly
{\em thermodynamic equilibrium}.\cite{ketcham69,elbaum93,sazaki22} A possible
hypothesis is that the arrest of the contact line in the macroscopic experiments
is  the result of some yet not fully understood non-equilibrium process related
to the pinning of the contact line due to freezing.\cite{thievenaz20} Be as it
may, a striking observation of these experiments is the increase of contact
angle as the temperature of the ice surface decreases,\cite{demmenie25,sarlin25}
reaching an amazing value of   150$^{\circ}$  at 180~K.\cite{demmenie25} Further
work on wetting dynamics and capillarity is needed to understand this behavior and 
its relation to the low temperature ice surface.\cite{huerre25}


\section{Is the observation of surface steps consistent with ice premelting?}

\begin{figure*}
		\includegraphics[width=\textwidth,trim=0 2cm 0 0]{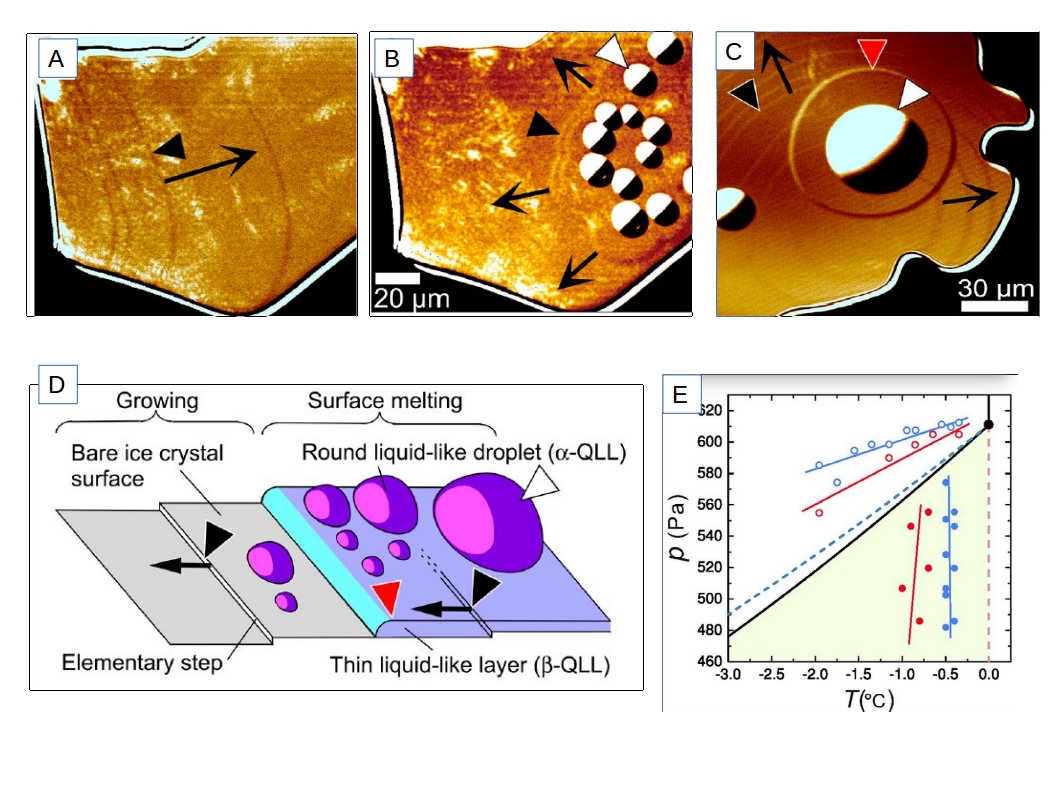}
\caption{Experimental observation of the ice basal face in the neighborhood of the triple point. A) At slight supersaturation, elementary steps appear and spread horizontally. B) At higher saturation, droplets condense on the surface. C) At yet higher saturation, thick premelting films can appear below the droplets. D) Sketch showing, from left to right, the sequence of events observed in A), B) and C), as interpreted in Ref.\onlinecite{sazaki12}. E) The ice phase diagram, illustrating the phase boundary lines in black, and the metastable prolongation of the condensation line in dashed black. The red line shows the region where condensation of water droplets first appears. The blue line shows the region where a second wetting film below the liquid droplets first appears. Notice the similarity with the sketch of Figure \ref{fig:phase_diagram} predicted from the theory described in Section IX. Figures A, B, C and D reproduced with permission from Sazaki et al.\cite{sazaki12} PNAS {\bf 109} 1052 (2012). 
Figure E reproduced with permission  from Murata et al.\cite{murata16} PNAS {\bf 113} E6741 (2016). 
	\label{fig:sazaki}}
\end{figure*}

From the above discussions, there appears to emerge  a consistent picture indicating that ice does not completely surface melt, but it does build up a significant premelting film attaining the scale of the nanometers at the triple point.

This view arises from the critical analysis of experimental measurements of
premelting film thickness,\cite{elbaum93,dosch95,bluhm02,sadtchenko02,mitsui19}
experimental observation of water
droplets,\cite{ketcham69,elbaum93,gonda99,sazaki13} computer
simulations,\cite{conde08,limmer14,qiu18,pickering18,llombart19,llombart20,llombart20b,berrens22} 
and theoretical considerations.\cite{benet19,llombart20,luengo22b,baran24b}

However, this interpretation is yet to meet an important challenge.

\change{In experiments performed by the Hokaido team, atomically smooth ice surfaces were prepared by epitaxial deposition from pure water vapor. This careful sample preparation, together with} advanced optical microscopy have allowed in the last
decade to observe (and report) the ice surface with unprecedented
detail.\cite{sazaki13,murata16,sazaki22}  Particularly, it was shown that  the basal ice surface at exceedingly  small supersaturation exhibits elementary steps in perfect agreement with the expected lattice structure, and results in crystal growth by terrace spreading or spiral growth.\cite{sazaki13,murata16,sazaki22} But the general wisdom in crystal growth theory is that this kind of {\em smooth} surfaces are in a highly ordered, low temperature state.\cite{safran94,chaikin95,feigelson04,akutsu15} The question then arises, how can one reconcile the step formation expected for a highly ordered surface, with the highly disordered state expected for a premelted surface? Sazaki and collaborators interpreted their own observation as evidence for the lack of premelting layers on the ice surface, most noticeably, in Ref.\cite{asakawa16}, but have thereafter sometimes stand  by,\cite{sazaki22} sometimes moderated this claim.\cite{murata19,nagata19}

However, a recent combination of simple models borrowed from crystal growth
theory and capillary wave theory, shows that the observation of significant
premelting and step formation are both mutually acceptable and reconcilable.\cite{benet16,benet19,sibley21}

\subsection{Healing distance of a surface} 

Before embarking into the details, however, it is worth digressing on the
healing distance, a concept that is important in the theory of wetting on
patterned substrates.\cite{degennes04} The question here is, whether a thin film
deposited on a patterned substrate will 'heal' the holes left by the pattern and
lead to a fully flat liquid-vapor surface, or, alternatively, whether it will
adapt to the substrate and lead to a liquid-vapor surface that reveals the shape
of the pattern. The decision  depends on a balance between the free energy  cost
of increasing the surface area by following the pattern; and the free energy
cost of healing the surface by changing the preferred equilibrium film
thickness. The former cost is governed by the liquid-vapor surface tension,
$\gamma_{\lv}$, while the latter is controlled by the interface potential,
$g(h_e)$. The result emerging from the theory of wetting is that there is a
relevant healing distance, or correlation length,
$\xi=\sqrt{\gamma_{lv}/g''(h_e)}$ (with $g''$ the second derivative of $g(h)$
with respect to $h$) which dictates the film's decision. If the surface tension
is very large, $\xi$ is large, and the film fills all holes with lateral
dimension smaller than the healing distance
to save the cost of creating extra surface area (as the film's height remains
correlated over length scales larger than the pattern relief); if, however, the
interface potential is strong, $\xi$ is small, and the liquid film will follow
the shape of the pattern, to save the free energy cost of perturbing the film thickness away from its equilibrium value (as the pattern relief is now of greater scale than the film height correlation length).

It perspires that, for liquid films with a healing distance smaller than the
typical distance between neighboring steps, one could envisage a premelting liquid film that would adapt to the steps. Optical observation of the ice surface would then exhibit traces of the steps  created on the buried solid surface, despite the presence of a highly disordered premelting film atop. 

Using our data for $g''$ reported in Fig.\ref{fig:disjoining} together with the surface tension
of liquid water, we find that $\xi\approx 1$~nm for the basal face approaching
$T_t$. This is somewhat larger (but in the order of magnitude) than the scale of
a step, ca. one molecular diameter, but much smaller than the typical distance
between steps observed in experiment. This suggests that, indeed, a premelting film could mimic  faithfully the shape of the solid ice underneath, so that the apparent conflicting observations of ice premelting and surface steps can be reconciled.

\subsection{Smooth and rough interfaces}

The discussion above is appealing, but it does not explain why and when does a crystal plane show steps and terraces as does the
basal plane of ice crystals, or whether such surface steps can be observed in a premelted interface. 

To understand this, it is required
to make the important distinction between smooth and rough crystal interfaces.\cite{rottman84,bienfait92,chaikin95,akutsu15} Smooth interfaces are those that are bound to the underlying
crystal lattice. They exhibit perfectly terminated
crystal planes, and show limited surface height fluctuations of the order of one or a few lattice spacings. On the other hand,
rough interfaces unbind from the underlying crystal lattice and exhibit very large surface fluctuations, very much as in fluid interfaces.
This apparently sophisticated distinction is
crucial for the behavior of a crystal.  In the absence of surface defects, smooth surfaces
can only grow by a slow activated process of two dimensional nucleation, where a small critical terrace is formed and then spreads horizontally along the surface.
On the other hand, rough surfaces can grow without an activation barrier in a continuous fashion and exhibit growth rates that are linear
in the saturation. As temperature increases, smooth surfaces gradually disorder and can become rough across a so called,
roughening transition. This transition was first conjectured in a famous paper by Burton, Cabrera and Frank in the late 1950's.\cite{burton51} But the
fine details  have not been understood until the late 1970's,\cite{vanbeijeren77,chui78,saito80} where it was fully characterized and shown to be a fluctuation dominated
phase transition of the Kosterlitz-Thoules type.\cite{bienfait92,chaikin95,akutsu15}


Since both roughening and premelting are surface disorder phenomena, the question then is, whether the roughening transition must necessarily occur before the appearance of significant surface premelting or not. If this were the case, the observation of steps on the basal plane would imply there can be no premelting indeed, as suggested by Sazaki et al.\cite{asakawa15,asakawa16,murata16} 
But simple considerations suggest   this need not be the case. Indeed, at the
ice-water interface, the basal plane is thought to remain smooth up to the
triple point. This expectation is supported by evidences of two dimensional nucleation of ice from its melt,\cite{furukawa21} the visual observation of elementary steps in vicinal planes,\cite{murata22} and the indication of layer-wise growth in molecular dynamics simulations.\cite{nada05,rozmanov12,mochizuki23}
Since the ice-water interface can be considered as a limiting case of a premelted surface with a film of infinite thickness, one then can expect intuitively  that a buried basal surface under a thin premelting film should also be able to remain smooth up to the triple point.


In order to substantiate this claim, it is required to take explicitly in consideration the complex structure of the premelted interface. This can be achieved by describing the premelted layer  in terms of an ice/water surface, $z_{\sl}({\bf x})$ and a water/vapor surface,  $z_{\lv}({\bf x})$, which  bound the layer in between the bulk ice and vapor phases at arbitrary points ${\bf x}$ in a flat horizontal reference plane parallel to the interface (Figure \ref{fig:snapshot}).\cite{benet16,benet19,llombart20b} 

The free energy cost of a given realization of the ice/water surface profile may
be assessed by means of the Sine-Gordon Hamiltonian, as:\cite{chaikin95}
\begin{equation} \label{eq:sinegordon}
\Delta \Omega_{\rm SG} =
\int d{\bf x} \left(
\frac{1}{2}\tilde\gamma_{\sl} (\nabla z_{\sl})^2 -u \cos(\frac{2\pi}{b}
z_{\sl}) \right)
\end{equation}
where the first term accounts for the free energy cost of increasing the surface
area, as dictated by the solid-liquid surface stiffness (for practical matters
the difference between a stiffness coefficient $\tilde \gamma$ and the
corresponding surface tension $\gamma$ can be ignored here), while the second
term  penalizes the deviations of the ice/water surface away from the underlying
lattice spacing, $b$; with $u$ a lattice strength parameter. 

For a simple solid-liquid interface, the distinction between rough and smooth interfaces can be made 
quantitatively by the calculation of the surface height fluctuations. But for practical matters, it proofs 
convenient to better consider the problem by transforming the surface profile, $z({\bf x})$ into Fourier space, 
as $z({\bf q})$, with ${\bf q}$ a wave-vector along the plane of the interface.
This is convenient, because it allows to obtain simple analytical results for the
Sine-Gordon Hamiltonian. Particularly, the mean squared surface fluctuations 
are given in Fourier space as:\cite{chaikin95,nelson04}
\begin{equation}\label{eq:hdeq}
    < \delta z_{\sl}^2(q) > = \frac{k_BT}{A} \frac{1}{ w + \tilde\gamma_{\sl} q^2}
\end{equation}
where $w\propto u$ is a renormalized lattice strength parameter
which sets how strongly is the solid surface tied to the underlying lattice. 
At low temperature, $w$ is finite, and the surface fluctuations $< z^2(q) >$
remain finite at all wave-vectors. However, as the temperature increases, $w$ is
renormalized by thermal fluctuations and eventually vanishes  at the roughening
temperature, $T_R\propto \tilde\gamma_{sl} b^2$,  irrespective of the cohesive energy
of the lattice. Above this temperature,  there is no longer a cutoff in the
denominator of Eq.(\ref{eq:hdeq}) and the surface height fluctuations diverge in
the limit of small wave-vectors. Thus, the distinction between finite or
diverging height fluctuations allows to identify whether a simple interface is rough or smooth.

For the more complex problem of a premelted interface, one needs to further
account for the surface fluctuations of the water/vapor surface bounding the
premelting film from the vapor phase. This can be achieved by using the
so-called Capillary-Wave Hamiltonian:\cite{evans92,rowlinson82b}
\begin{equation}
\Delta \Omega_{\rm CW} =
\int d{\bf x} \left(
\frac{1}{2}\gamma_{\lv} (\nabla z_{\sl})^2 + g(h) \right)
\label{eq:capillary}
\end{equation}
where, again, the first term accounts for the free energy cost of increasing the
water/vapor surface area, while the second term, with $g(h)$ the interface
potential, penalizes excursions of the premelting film thickness away from the
preferred equilibrium value (as dictated by Eq.(\ref{eq:derjaguin})).

The total surface free energy of the premelted interface is obtained as a sum
of the  Sine-Gordon and Capillary-Wave Hamiltonians,
Eq.(\ref{eq:sinegordon},\ref{eq:capillary}), which are coupled via the premelting film
thickness, $h({\bf x})=z_{\lv}({\bf x})-z_{\sl}({\bf x})$

This coupled SG+CW model of surface fluctuations can be solved approximately,\cite{benet16,benet19,llombart20b} 
using an elegant renormalization method due to Saito (notice this approach was erroneously
attributed to Safran in the original Ref.\onlinecite{benet16}).\cite{saito80}  In this
case, the outcome is an expression for the surface height fluctuations of the
ice/water and water/vapor surfaces, as well as the corresponding cross correlations.
 \begin{equation}\label{eq:h2surf}
 \begin{array}{ccc}
 \langle |z_{\sl}^2({\bf q})| \rangle & = &
 \displaystyle{     \frac{k_BT}{A} \frac{g'' + \gamma_{\lv} q^2}{
 		w g'' + (g''\Sigma + w \gamma_{\lv})q^2 + \gamma^2 q^4
 	} } \\
 	& & \\
 	\langle |z_{\lv}^2({\bf q})| \rangle & = &
 	\displaystyle{      \frac{k_BT}{A} \frac{w + g'' + \tilde\gamma_{\sl} q^2}
 		{
 			w g'' + (g''\Sigma + w \gamma_{\lv})q^2 + \gamma^2 q^4
 		}  } \\
 		& & \\
 		\langle z_{\sl}({\bf q}) z_{\lv}^*({\bf q}) \rangle & = &
 		\displaystyle{      \frac{k_BT}{A} \frac{g''}
 			{
 				w g'' + (g''\Sigma + w \gamma_{\lv})q^2 + \gamma^2 q^4
 			} 
 		}
 		\end{array}
\end{equation} 		
where $\Sigma =\gamma_{\sl}+\gamma_{\lv}$,
$\gamma^2=\gamma_{\sl}\,\gamma_{\lv}$ and $g''=d^2 g/ dh^2$.

Despite the increased algebraic complexity, the overall behavior of these
compound interface is very similar to that of the isolated ice-water interface
described in Eq.(\ref{eq:hdeq}). Indeed, for finite $w$, it is seen that all
surface fluctuations remain finite in the limit of $q\to 0$. However, when $w$
vanishes, all three surface fluctuations diverge in the limit of small
wave-lengths. The difference is that, now, the roughening transition occurs for
$T_R \propto \Sigma b^2$, \cite{benet16} which is significantly higher than the roughening transition at the ice-water interface (as anticipated in Ref.\cite{mikheev91} from purely qualitative arguments). This confirms that 1) the premelted basal interface can remain smooth all the way up to the triple point and 2)  there is no conflict between the observation of surface steps on the basal face and the existence of substantial ice premelting.

\subsection{Digression: Can ice crystals be overheated?}

Most crystalline solids are difficult to overheat, because they start melting instantaneously at their surface.\cite{nozieres92,dash95} However,  it has been suggested that the observation of water droplets on the ice surface could imply the possibility to overheat ice.\cite{knight71} The rationale is that melting at the surface can only proceed by increasing the equilibrium premelting film thickness, which, in the absence of complete surface melting lies in a free energy minimum (c.f. Figure \ref{fig:gminmod}) and is thus reluctant to move away from the minimum.\cite{baran24b} In practice, however, the free energy minimum is very shallow, as discussed before. However, smooth interfaces are known to grow epitaxially by activated nucleation, and whence, are also expected to have an activation barrier for the epitaxial layers to melt.\cite{weeks77} This provides an additional reason for overheating. Indeed, Elbaum et al. reported that the basal ice surface could survive overheating of up to 0.2 to 0.3~$^{\circ}$C,\cite{elbaum93}, while  droplets on the basal ice surface  at very nearly the melting point have been reported in work by Gonda and Sei.\cite{gonda99} In computer simulations, perfect ice mono-crystals exposed only at the basal surface could be superheated up to 2$^{\circ}$C, but similar behavior for the prism face was not found.\cite{baran24b}

\section{The equilibrium structure  of premelting films}

\begin{figure*}
	\includegraphics[width=0.9\textwidth]{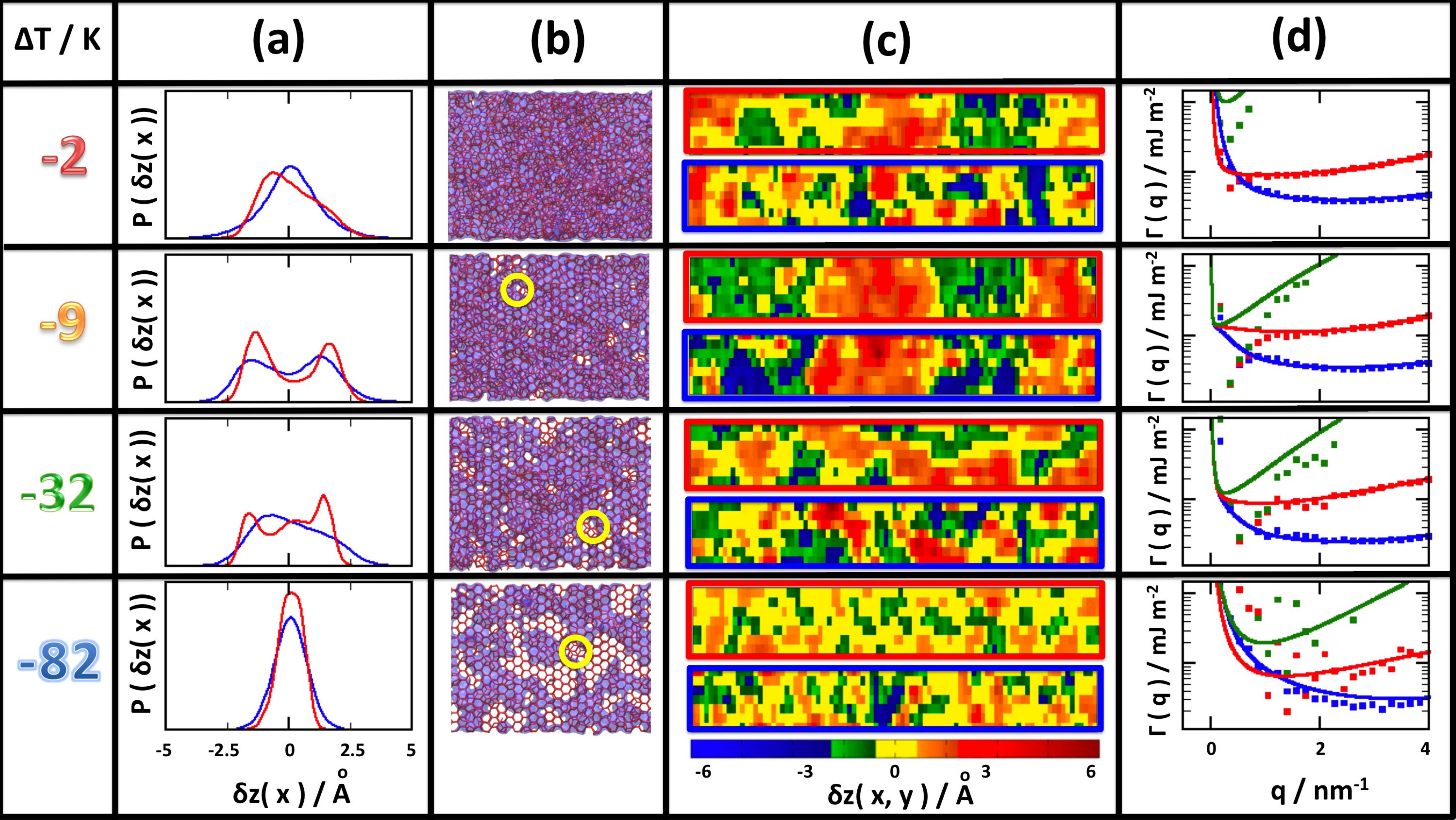}
	\caption{Equilibrium structure of the ice basal face as described by the TIP4P/Ice model. (a) The probability density of film heights reveals two surface phase transitions as shown by the transformation from unimodal to bimodal and back to unimodal distributions as temperature increases. (b) snapshots of surface coverage of liquid-like molecules on top of the bulk ice. (c) Heat map of height distributions shows the emergence of thick correlated up and down domains at intermediate temperatures. (d) The spectrum of surface height fluctuations can be fitted to the an analytical model of Eq.(\ref{eq:h2surf}) in order to estimate step free energies (c.f. Fig.\ref{fig:stepfree}). Reproduced from Llombart et al.\cite{llombart20b} Sci. Adv. {\bf 6} eaay9322 (2020). Distributed under a Creative Commons Attribution License CC-BY.
		\label{fig:basalstruc}
	}
\end{figure*}

From the previous sections, it becomes clear that the ice-vapor interface has a complex three-dimensional structure featuring a significant layer of disordered water on top of bulk ice. Therefore, reducing the problem to the characterization of a single dividing surface as is usual in surface thermodynamics misses out plenty of interesting details. 

A more detailed and insightful characterization may be achieved by using the
ice/water and water/vapor surface profiles, $z_{\sl}({\bf x})$, $z_{\lv}({\bf
x})$, discussed above.  Using these definitions, it is possible to monitor the deviations of the local surface height, $\delta z({\bf x})$ from the instantaneous average position $\overline{z}$, for either the ice/water, $\delta z_{\sl}({\bf x}) = z_{\sl}({\bf x}) - \overline{z}_{\sl}$, or water/vapor surfaces, $\delta z_{\lv}({\bf x}) = z_{\lv}({\bf x}) - \overline{z}_{\lv}$. The identification of this rather sophisticated order parameters turns out to be extremely useful, and allows to identify a number of structural transformations on the ice surface.

As an example, Fig.\ref{fig:basalstruc}-(a) shows the probability distribution  of $\delta z_{\sl}({\bf x})$ and $\delta z_{\lv}({\bf x})$ as temperature increases from $\Delta T = T - T_t = -82$~K, to $\Delta T = -2$~K on the equilibrium basal surface of ice as described by the TIP4P/Ice model. At the lowest temperature, the distribution of surface heights is a narrow Gaussian, as expected for a sharp low temperature interface. In the language of the statistical thermodynamics of soft solid interfaces,\cite{rommelse87,prestipino95,weichman96,woodraska97b,jagla99}
this state is designed as an Ordered Flat phase (OF). As temperature increases,
steps and terraces start to proliferate on the ice surface, as revealed by the
multi-modal character of the probability distribution. A surface structure with
proliferation of terraces but limited surface height fluctuations is known as a
Disordered Flat Phase (DOF), as first identified by Rommelse and den Nijs in simple
column models of the solid interface.\cite{rommelse87} Finally, at yet higher
temperature, the multi-modal probability distribution vanishes, and now becomes a
broad Gaussian distribution. This transformation occurs concurrently with a
significant increase of the premelting thickness, so it appears that the
interface trades the entropy of step proliferation in the DOF phase, for the
translational entropy of the liquid like water molecules. This last surface
phase was named as a High-Temperature Reconstructed phase (HT-RF) in
Ref.\onlinecite{llombart20b}, although the name of Flat Premelted phase (FP) would have been 
more appropriate. 

The distinction between OF and HT-RF phases, with unimodal distributions, and  the DOF phase, with a multi-modal distribution of surface heights, can be seen in Fig.\ref{fig:basalstruc}-(c), which displays a heat-map of the instantaneous surface height fluctuations. Corresponding to the DOF phase, the basal surface becomes heterogeneous, and correlated up and down domains of the surface height are clearly visible.  A similar heterogeneous distribution of surface heights was also observed in studies of the mW model for the basal face at low temperatures.\cite{pickering18,qiu18}

The study of the prism facet reveals also the existence of three different
surface phases, albeit with the phase transitions occurring at  significantly
different temperatures from those observed for the basal face.\cite{llombart20b} 
A related study for the TIP4P/2005 water model studied the height-height surface
correlations in real space and also found evidence of the existence of three
different phases on basal and prism faces of ice.\cite{berrens22}

The characterization of different surface phases at the ice-vapor interface is extremely appealing, as it could provide for the explanation of the Nakaya diagram. Indeed, one expects that the crystal growth rates could change significantly across the surface phase transitions, resulting in non-monotonous behavior. Without any other factual evidence, in 1982 Kuroda and Lacmann made the bold hypothesis that the ice surface should exhibit three different surface phases.  They conjectured that a mismatch in the transition temperatures in prism and basal facets could lead to the alternating growth rates that are required to explain the Nakaya diagram.\cite{kuroda82}  Four decades after the formulation of the Kuroda-Lacmann theory, computer simulations appear to support that hypothesis.\cite{llombart20b}

This conjecture seems reasonable and appealing, but it remains to be substantiated with actual predictions of the crystal growth rates.  This begs the question 1) How representative is the equilibrium surface structure during the out of equilibrium growth process of ice in the atmosphere? 2) How could the crystal growth rates be calculated from the equilibrium surface properties?

\section{Premelting films away from the sublimation line}

\label{sec:noneq}

In practice, the ice-vapor interface can only be realized in equilibrium
conditions for state points along the sublimation line,
a situation which is actually more the exception than the rule. \change{Indeed},
ice samples in the atmosphere will be most often found in conditions away from
equilibrium.\cite{libbrecht22}  In this situation, the interface does no longer
remain uniform, but will display  pattern formation,  inhomogeneities and
instabilities.\cite{misbah10} However, contrary to some simple situations, the
departure from equilibrium cannot be fully interpreted just as a measure of the
distance of the vapor away from the sublimation line. The vicinity of the (metastable)
liquid-vapor coexistence line means that, together with the
sublimation/desublimation processes, it is necessary to account for the
competing evaporation/condensation process of the marginally metastable liquid
phase (c.f. Figure \ref{fig:phase_diagram}).

A simple manifestation of departure from equilibrium that was already mentioned
is the spontaneous formation of water droplets by vapor deposition on the ice
surface.\cite{ketcham69,elbaum93,gonda99,sazaki12} According to the Laplace equation, this phenomenon can only occur when the pressure inside the liquid phase is larger than that in the surrounding vapor phase.\cite{rowlinson82b} In material equilibrium, with both phases at equal chemical potential, this occurs as soon as the vapor is saturated over water.

However, the experiments performed by the Hokaido team show that water droplets
on the ice surface form only in conditions of significant supersaturation over
water vapor, when
the pressure is increased beyond a temperature dependent threshold
 that lies well above the condensation line (c.f. Fig.\ref{fig:sazaki}). \cite{asakawa16,murata16}

In order to understand this behavior, it is convenient to extend the free energy model of premelting films, Eq.(\ref{eq:sfeh}), which is only valid along the coexistence line, to the general case of arbitrary pressure of the vapor phase, $p_{\vap}$. i.e.:\cite{luengo21}
\begin{equation}\label{eq:wzz}
 \Delta \omega(h) =  \gamma_{\sl} + \gamma_{\lv} + g(h) - (p_{\sol}-p_{\vap})  z_{\sl} - (p_{\liq} - p_{\vap}) z_{\lv}
\end{equation}
where  $p_{\sol}$ and $p_{\liq}$ are the pressures of the corresponding bulk phases at equal temperature and chemical potential as the mother vapor phase; while 
$z_{\sl}$ and $z_{\lv}$ are the heights of the  ice/water and water/vapor surfaces bounding the premelting liquid film from below and above, respectively, with the premelting film thickness now given as $h=z_{\lv}-z_{\sl}$.

This is a simplified free energy model that accounts for a uniform out of equilibrium solid/vapor interface. Borrowing tools from fluctuation theory and near equilibrium physics, this free energy may be used to describe the dissipative dynamics of the interface.\cite{goldenfeld92} The key issue here is that the dynamics of a system close to equilibrium is dictated by deviations of the free energy from its equilibrium value. Particularly, it is expected that the time derivative of the relevant variables is linear in the free energy gradients. Therefore,  the growth rate at the ice/water surface bounding the premelting film from below, which results from freezing/melting events, may be qualitatively described as $r_{\sl} \propto - k_{\sl} \partial \Delta \omega /\partial z_{\sl}$, with $k_{\sl}$ the kinetic growth coefficient. In analogy, the growth rate at the water/vapor surface bounding the film from above, which results  from condensation/evaporation events, may be described as $r_{\lv} \propto - k_{\lv} \partial  \Delta \omega /\partial z_{\lv}$, with $k_{\lv}$ the corresponding rate coefficient.\cite{sibley21}
The evolution of the solid/liquid bounding surface will be dictated by the
occurrence of freezing/melting events, and so $d z_{\sl}/dt \propto r_{\sl}$. On the other hand, the evolution of the liquid/vapor surface is  mainly given as a result of condensation/evaporation events, but also depends on  the freezing/melting events of the lower surface, due to the difference in density of ice and water; whence  $d z_{\lv} / dt \propto r_{\lv} - r_{\sl}$.

Using this set of phenomenological equations, it is possible to show  there exists a   {\em quasi-stationary} regime where ice can gently grow ($d z_{\sl}/dt > 0$), or melt ($d z_{\sl}/dt < 0$) while the premelting film remains in a steady state of constant film thickness, ($d h/dt = 0$). It corresponds to a physical situation where the lower surface of the premelting film freezes at the same rate as water vapor condenses on the upper surface. This is an extremely relevant observation that was already recognized by Kuroda and Lacmann,\cite{kuroda82,kuroda83,kuroda90} and discussed qualitatively in computer simulation studies.\cite{pfalzgraf11,neshyba16,mohandesi18,pickering18,shi25} A modern rewrite in the language of wetting physics provides the condition for the quasi-stationary
regime in striking resemblance with the equilibrium condition of premelting
films, Eq.(\ref{eq:derjaguin}):\cite{sibley21} 
\begin{equation}\label{eq:kderjaguin}
      \Pi(h_{\rm qs}) = - \Delta p_k
\end{equation}
where $h_{\rm qs}$ stands here for the steady state film thickness in the
quasi-stationary regime, while the kinetic Laplace pressure, $\Delta p_k$ is the out-of-equilibrium analog of $\Delta p$ in Eq.(\ref{eq:derjaguin}): 
\begin{equation}\label{eq:defpk}
\Delta p_k = \frac{\rho_{\liq} k_{\lv}}{\rho_{\sol}k_{\sl}+\rho_{\liq}k_{\lv}}
[p_{\liq}-p_{\vap}] -
\frac{\rho_{\sol} k_{\sl}}{\rho_{\sol}k_{\sl}+\rho_{\liq}k_{\lv}}
[p_{\sol}-p_{\liq}] 
\end{equation}
The simple model described above provides a valuable framework for the understanding of growth morphology of premelting films. First of all, it shows that the notion of a well defined premelting film remains meaningful in out of equilibrium conditions.\cite{kuroda90,lowen91} Second,
it shows that the corresponding steady state premelting film thickness can be measured from knowledge of the equilibrium disjoining pressure and the kinetic coefficients of the growth process. And third, it shows that the expected wetting behavior of the quasi-stationary regime can be described in analogy to that expected for a wetting film on an inert substrate.\cite{lowen91}

Particularly, it is known that the condition for mechanical equilibrium of a
liquid droplet on an inert substrate is a positive Laplace pressure, $\Delta p >
0$. So, likewise, in the quasi-stationary regime, with a living substrate
feeding from the premelting film, the condition of mechanical equilibrium of the
droplet now becomes $\Delta p_k > 0$.\cite{sibley21} Solving for the marginal
condition $\Delta p_k(p_v,T) = 0$  provides a {\em kinetic condensation line},
which dictates the pressure at which a droplet can first start growing atop the
liquid/vapor surface \change{(c.f. dashed blue line of Fig.\ref{fig:phase_diagram})} . This can be worked out easily from the above equation  and leads to a kinetic phase boundary line of Clausius-Clapeyron type:
\begin{equation}
 \ln p_{kc}/p_t  =  \frac{\Delta H_{kc}}{R} \left ( \frac{1}{T_t} - \frac{1}{T} \right)
\end{equation} 
where $p_{kc}$ stand for the {\em kinetic condensation pressure}, and $\Delta
H_{kc}>\Delta H_{\lv}$ is an effective enthalpy of vaporization. This is always larger than the equilibrium vaporization enthalpy, $\Delta H_{\lv}$, so that  the condition for a droplet condensing on top of the growing premelted ice surface lies always above the equilibrium condensation line, in full agreement with the observations of the Hokaido team \change{(c.f. red line of Fig.\ref{fig:sazaki}.E)}.\cite{asakawa16,murata16}

Another relevant phase line corresponds to the solution of Eq.(\ref{eq:kderjaguin}) in the special case that 
the steady state thickness $h_{\rm qs}$ attains the value corresponding to the
wetting spinodal, i.e., where the second derivative of the interface potential
$g(h)$ first vanishes. The kinetic spinodal line  lies above the kinetic
coexistence line, and describes the phase points where a quasi-stationary regime
can no longer hold \change{(c.f. dashed red line of Fig.\ref{fig:phase_diagram})}. In this situation, the freezing rate at the ice/water
surface can no longer match the condensation rate at the water/vapor surface, so
that the premelting film thickness starts growing without bounds. This
corresponds to a situation of {\em kinetic} complete surface melting, where the  ice sample is freezing underneath the liquid film, but the liquid film thickness is diverging due to vapor condensation at a faster pace.\cite{sibley21}

The kinetic phase boundaries can be extended to a temperature higher than $T_t$,
where they then describe the evaporative regime of the ice crystal.  In this
case, the kinetic spinodal line dictates the conditions where the evaporation at
the water/vapor surface can no longer match the rate of melting at the ice/water
surface. The crystal here will exhibit wet evaporation, and actually disappear
from a puddle of water. In between this line and the liquid-vapor coexistence line, there lies a prolongation of the kinetic condensation line, which bounds the regions of thermodynamic variables where water droplets can form as a result of ice melting.
This rather unexpected behavior was reported by Asakawa and collaborators, as a
set of  boundary lines running almost parallel to the melting line.
Intriguingly, the lines measured in Ref.\cite{murata16} where slightly below
$T_t$, rather than above $T_t$ as predicted in the kinetic model described here.
A shift of the triple point due to the presence of impurities could help explain
the discrepancy with respect to the theoretical expectation.

The different growth regimes discussed in this section are sketched in qualitative manner in Figure~\change{\ref{fig:phase_diagram}, to be compared with the experimental results of Fig.~\ref{fig:sazaki}.E}. However, an important thing to notice is that the
quasi-stationary growth regime described here, and the corresponding kinetic boundary lines, are not a universal feature of the substance. As recognized by  Kuroda and Gonda long time ago,\cite{kuroda90} they crucially depend on the growth mechanism followed by the crystal. Depending on the crystal plane, initial conditions and sample history, this can be nucleated, spiral or linear growth. The linear growth discussed above, however, is the one leading to the simplest algebraic results and is used here for the sake of illustration.

In practice, in out of equilibrium conditions there occur film instabilities, and the mean field free energy model of Eq.(\ref{eq:wzz}), corresponding to films of uniform thickness is not sufficient to describe the complex out of equilibrium dynamic that actually takes place. However, accounting for inhomogeneities using the SG-CW free energy model of Section VII.B allows to describe complex  non-equilibrium phenomena in remarkable agreement with experimental findings (c.f.: Figure \ref{fig:mesodrop}).\cite{sibley21}

Be as it may, the kinetic phase diagram that emerges from this theoretical treatment appears to be extremely relevant for the understanding of crystal growth in the atmosphere.
Particularly, the formation of complex crystal habits displayed in the Nakaya diagram occur somewhat above the equilibrium condensation line.\cite{harrington24} It is very tempting to speculate that this threshold of complex habit behavior is related to the
first emergence of water vapor condensation effects on the crystal surface as described by the Kinetic condensation line.\cite{sibley21} Confirmation of this hypothesis would have a significant impact in the understanding of crystal growth from the vapor.


\section{Step free energies and the Nakaya diagram}

From the above discussion it follows that in a narrow region of supersaturation in between the sublimation line and the kinetic condensation line which is relevant for snowflake growth in the atmosphere,\cite{libbrecht22} ice grows gently under a steady state regime of constant premelting thickness.\cite{kuroda82} This corresponds to the linear response regime, where the growth rates are dictated by the equilibrium surface structure, as discussed in Fig.\ref{fig:basalstruc}. Indeed, in linear response theory, the out of equilibrium dynamics of a system is fully dictated by the correlated dynamics of the system in equilibrium.\cite{hansen86,goldenfeld92} 

The fluctuation formula of Eq.(\ref{eq:h2surf}) then become very valuable.
Fitting these equations to the spectrum of surface fluctuations as obtained from
computer simulations (Fig.\ref{fig:basalstruc}-d) allows to assess the value of the lattice strength parameter, $w$. Not surprisingly, this, together with the surface stiffness coefficients, sets the free energy cost of creating an elementary step of height $b$ on the crystals' surface.

 Using results from the sine-Gordon model of  crystalline interfaces,\cite{bennett81,chaikin95} the free energy cost to create a step on the ice/water surface  amounts to 
$\beta_{\sl}=\frac{2b^2}{\pi^2}\sqrt{\gamma_{\sl}w}$.\cite{nozieres87} 
In the presence of premelting it is also required to account for the cost of
rising the water/vapor surface above the step so formed. By analogy with the previous result, it is expected that this free energy cost will be given by $\beta_{\lv} = \frac{2b^2}{\pi^2}\sqrt{\gamma_{\lv}g''}$.\cite{llombart20b} 
The total free energy cost of creating a step on the premelted surface must then be given as $\beta = \beta_{\sl} + \beta_{\lv}$.\cite{llombart20b} 

The rationale of these two equations may be understood as follows: the deviation
of a surface from planarity bears a free energy cost related to the increase of
surface area, which can be measured as $\gamma (dz/dx)^2 dx$, with the squared
derivative of  $z(x)$, used as a measure of the surface area increment.\cite{nozieres87,chaikin95} On average, the surface height increases by an amount equal to the full step height $dz=b$  over a horizontal distance on the  scale of the surface correlation length,   $dx=\xi$. This is  $\xi_{\sl}=\sqrt{\gamma_{\sl}/w}$ or  $\xi_{\lv}=\sqrt{\gamma/g''}$ for either the ice/water and water/vapor surfaces, respectively.  Then, the calculation of $\gamma (b/\xi)^2 \xi$, leads to the previous results up to numerical factors.

\begin{figure}
	\includegraphics[width=0.5\textwidth]{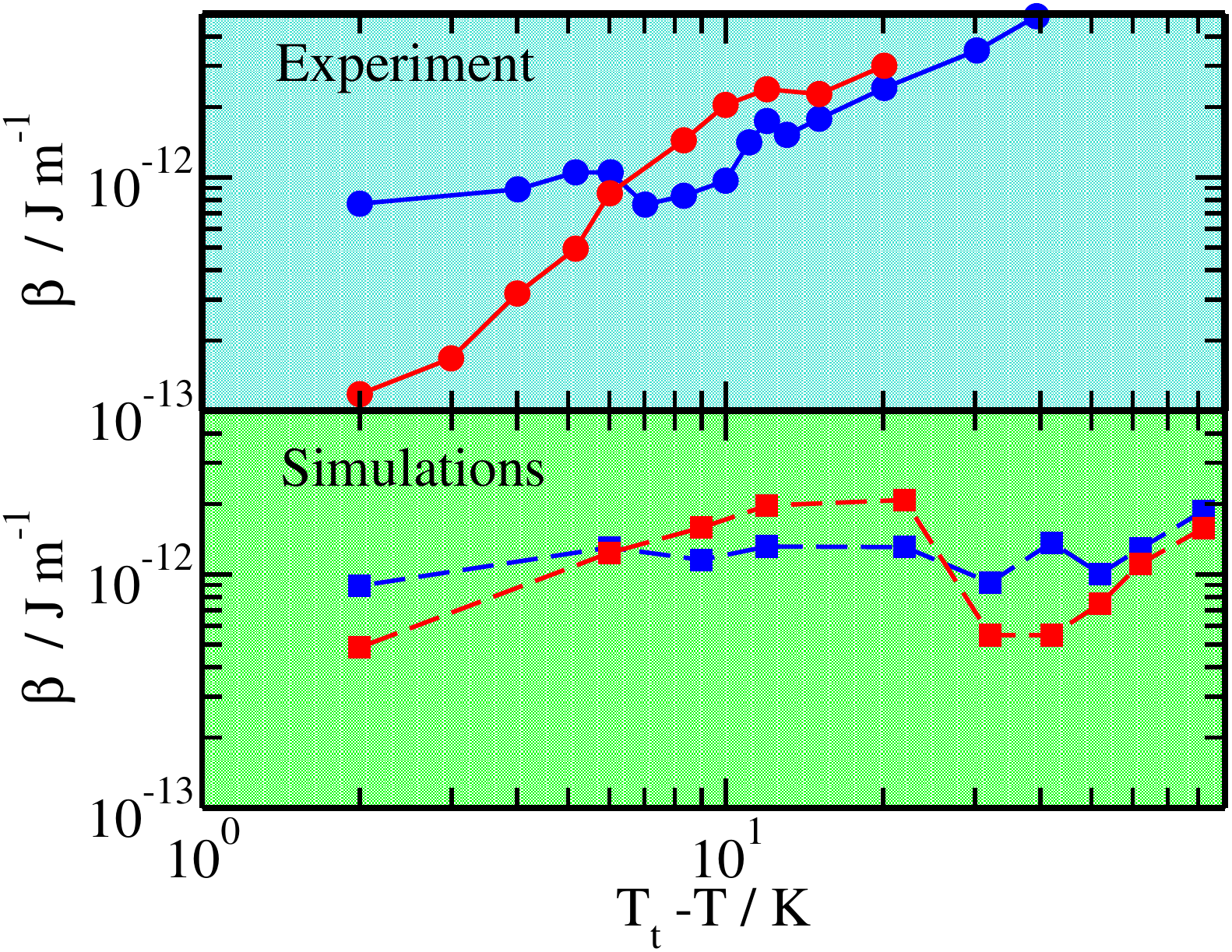}
	\caption{Step free energies of basal (blue) and primary prism (red) faces. The top panel shows experimental results as reported in Ref.\cite{libbrecht17}. The bottom panel shows simulation results for the TIP4P/Ice. In both cases, the step free energies alternate sequentially between the basal and prism faces as expected for the plate-column-plate transitions observed in the Nakaya diagram. Reproduced from Llombart et al.\cite{llombart20b} Sci. Adv. {\bf 6} eaay9322 (2020). Distributed under a Creative Commons Attribution License CC-BY. 
		\label{fig:stepfree}}
\end{figure}

Fig.\ref{fig:stepfree} compares the step free energies of prism and basal facets of ice crystals obtained from computer simulations for the  TIP4P/Ice model.\cite{llombart20b} The results are in order of magnitude agreement with measurements obtained directly from the growth of snow crystals,\cite{libbrecht17,libbrecht22} and show a non-monotonous behavior as required to explain the Nakaya diagram. Indeed, the step free energies appear higher or lower for either the basal or prism facets in alternating order.

Crystal growth in the atmosphere is thought to  occur by 2d nucleation,\cite{frank82,nelson98,libbrecht12,libbrecht22} but possibly also as a result of dislocations at the edge of the crystals.\cite{sei89,asakawa15,asakawa16,harrington24} In either case, the growth rates are inversely related to the step free energies.
According to the results of Fig.\ref{fig:stepfree},   in the close neighborhood of the triple point the prism face has lower step free energies and grows faster than the basal face, leading to formation of plate like crystal habits.  As temperature decreases, the step free energy of the prism face becomes larger than that of the basal face, the growth rate of the basal face takes over, and crystals adopt a columnar habit. Eventually, a new crossover between step free energies takes place, and crystals adopt again a plate like habit.  This scenario as explained from the measurement of step free energies in equilibrium conditions has been confirmed for the same water model by measurement of the relaxation dynamics of thick liquid films adsorbed on ice.\cite{shi25}

The alternating behavior of crystal growth rates of ice from the vapor has been reported in a number of experiments over the years,\cite{kobayashi67,lamb72,sei89,libbrecht12}  but a molecular explanation of this unusual behavior has remained elusive. However, thanks to the atomic scale resolution of computer simulations, the development of reliable force fields for water,\cite{abascal05,abascal05b} and the formulation of a suitable theoretical framework for the analysis of computer simulations,\cite{benet16,benet19,llombart20b} a consistent explanation appears to emerge. The 
ice surface structure is extremely complex, and cannot be explained in terms of the traditional crystal surface models based on the surface diffusion of isolated ad-atoms
that is adequate for the description of high energy solids very far away from
their triple point. Contrary to this venerable picture, the ice surface exhibits
all the way from 200~K to its melting point a significant amount of self-adsorption which increases with temperature and reaches the nanometer thickness in the neighborhood of the triple point. In this process, the basal and prism surfaces undergo a number of structural rearrangements,\cite{llombart20,berrens22} with profound impact on the crystal growth dynamics. The result of these transformations is that the growth rates become higher or lower for either the basal or prism facets in alternating order. Whereas the detailed structure of the surface phases is different, the underlying reason for the alternating growth rates is very much as anticipated by Kuroda and Lacmann many years ago.\cite{kuroda82}

\begin{figure*}[t]
	\centering
	\begin{minipage}[b]{0.33\textwidth}
		\begin{overpic}[width=\textwidth,trim=0 0 1cm 0,clip]{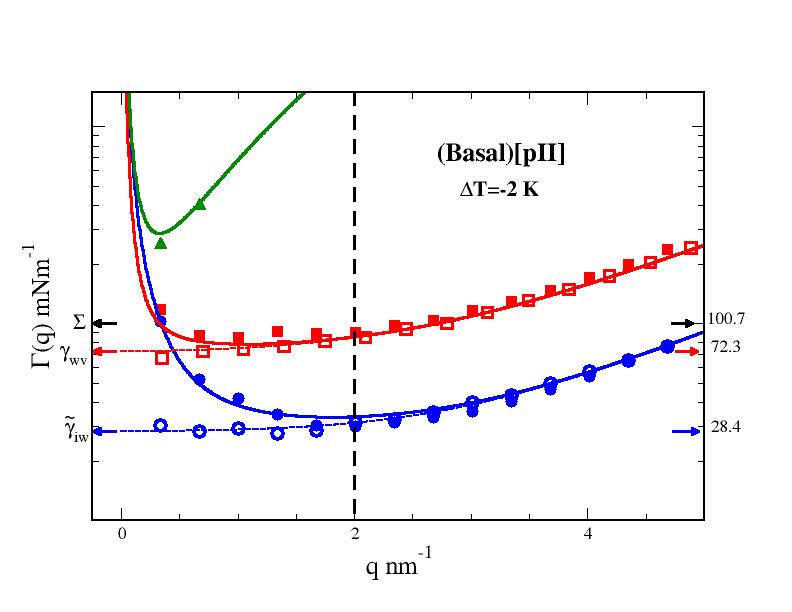}
			\put(2,72){\textbf{(a)}} 
		\end{overpic}
	\end{minipage}%
	\hfill
	\begin{minipage}[b]{0.33\textwidth}
		\begin{overpic}[width=\textwidth,trim=0 0 1cm 0,clip]{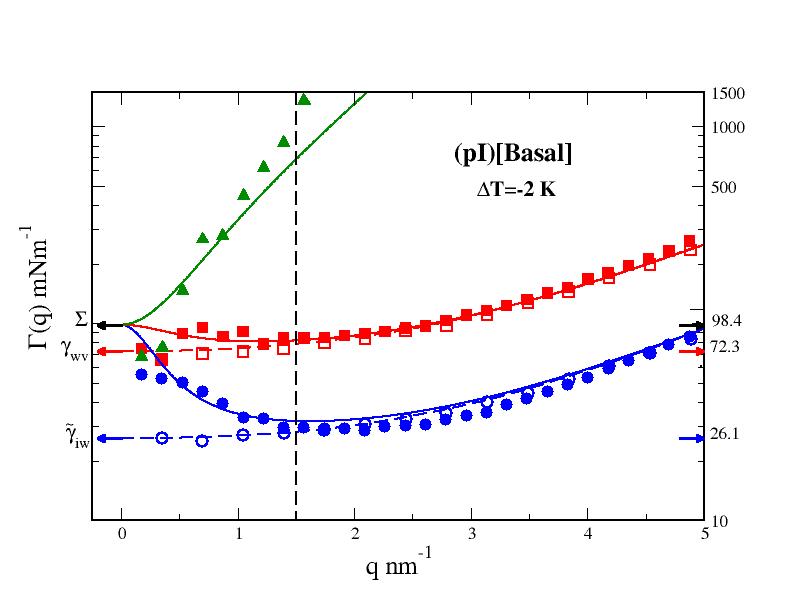}
			\put(2,72){\textbf{(b)}} 
		\end{overpic}
	\end{minipage}%
	\hfill
	\begin{minipage}[b]{0.33\textwidth}
		\begin{overpic}[width=\textwidth]{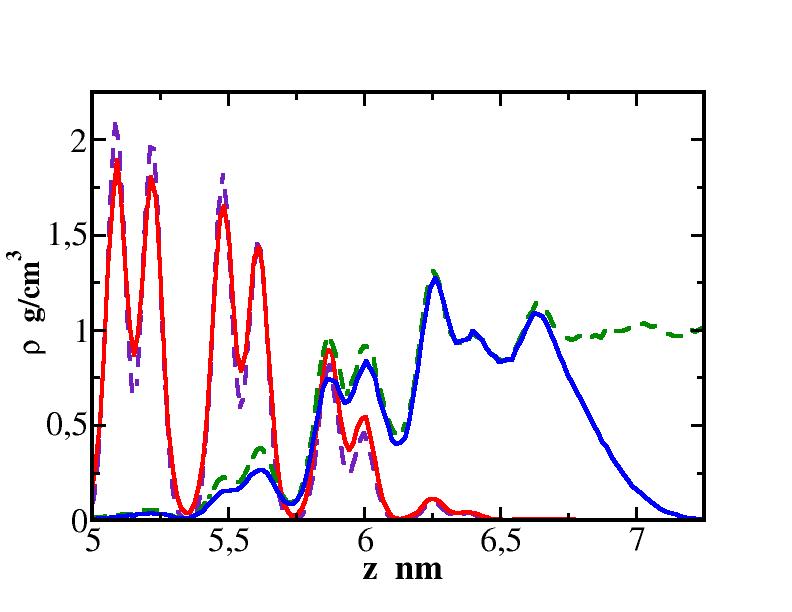}
			\put(2,72){\textbf{(c)}} 
		\end{overpic}
	\end{minipage}
		\caption{Interfacial structure and surface fluctuations of ice close
		to the triple point. (a) and (b) show the spectrum of surface
	   fluctuations for the basal and prism faces in the near vicinity of the
	triple point as described by the TIP4P/2005 model of water. Results are
   displayed as wave-vector dependent surface tensions of the ice/water (blue)
and  water/vapor (red) surfaces, together with the coupled correlations (green).
Symbols are results from computer simulations, while lines represent fits to the model of Eq.{\ref{eq:h2surf}}. The corresponding surface tensions for interfaces of bulk water with either bulk ice or bulk vapor are shown as empty symbols and look equal to those of the premelting film up to $q=2$~nm$^{-1}$, as indicated by the black dashed line. (c) Figure compares density profiles of the premelting film (continuous lines) with those of  the ice/water and water/vapor interfaces (dashed lines). The red and blue full lines are density profiles of solid-like and liquid-like molecules of the premelting film, respectively. The violet and green dashed lines are densities of solid-like  and liquid-like molecules from the ice/water and water/vapor interfaces, respectively. Reproduced with permission from from Benet et al.\cite{benet16}, Phys. Rev. Let. {\bf 117} 096101 (2016). Copyright \copyright 2016 American Physical Society. \label{fig:benet16} }
\end{figure*}

\section{How liquid is quasi-liquid?}

\label{sec:ql}

In wetting physics emphasis is made on the condensation of liquid films on a substrate.  Of course, it is understood that as long as the thickness of the film remains finite, the film is not strictly a bulk liquid, but there never is a need to emphasize this in the theoretical framework, whether condensation occurs in the form of a liquid film or a liquid droplet.

In the study of ice premelting, one assumes that the premelting film has its origin in the melting of ice like molecules. But as far as a thermodynamic description of the equilibrium ice-vapor interface is required, there is no need what so ever to consider whether the film is formed by ice melting or vapor deposition. In fact, from a dynamical point of view, the rates of adsorption/desorption and freezing/melting need to be exactly equal at equilibrium.

However, it has been traditional in the study of the ice-vapor interface to refer to the surface disordered layer as the {\em quasi-liquid} layer. This is fine, but it puts the emphasis of the problem on how different this could be from the expected bulk liquid, suggesting enigmatic properties and an aura of mystery 

Yet, if human kind had to give up computer simulations and be left with one single qualitative take away message from now more than 50 years of practice, this would be that macroscopic constitutive laws can be qualitatively trusted down to the nanometer scale. Within my experience this statement holds for all molecular liquids,  and the quasi-liquid layer is found not to be an exception.

Within a few Kelvin of the triple point, most computer
simulations,\cite{conde08,limmer14,qiu18,pickering18,llombart19,llombart20,llombart20b,berrens22} and experimental probes,\cite{elbaum93,bluhm02,sadtchenko02,mitsui19} indicate that the thickness of the premelting film has reached that scale. Based on this thumb rule, the quasi-liquid layer at the triple point is indeed very much like liquid water, despite its limited thickness.

A first evidence comes from the study of droplets. At the triple point, the ice-vapor surface tension differs from the sum of ice-water (ca. 30~mN$/$m, from Ref.\cite{fletcher70}) and water-vapor surface tensions (ca. 75.7~mN$/$m, from Ref.\cite{pruppacher10}) by $g(h_e)$, the minimum of the interface potential. From  Eq.(\ref{eq:cag}), this can be equated to $g(h_e)\approx-\frac{1}{2}\gamma_{\lv}\theta^2$. Using the upper bound of contact angles observed in experiment, the correction cannot possibly exceed $\sim -2$mN$/$m, 
which is actually less than the uncertainty in current estimates of the
ice-water surface tension, and
no more than 2~\% of the  result expected in the case of complete wetting., ca.
$75+30=105$~mN/J. So,  whatever the quasi-liquid layer is, it is sufficient to
exhibit surface tensions that do not differ by more than 2~\% from the value
expected for bulk liquid water at the interface of either ice or water vapor. 

Another evidence comes from the study of surface fluctuations of the premelting
film as studied from computer simulations.\cite{benet16} Using an adequate order
parameter,\cite{lechner08,nguyen15} it is possible to resolve the surface
fluctuations of the ice/film and film/vapor surfaces that bound the premelting
film from the bulk phases and measure their corresponding  wave-vector dependent
surface tension . These can then be compared with   the wave-vector dependent
surface tensions of independent simulations for the ice-water and water-vapor
interfaces.  The results of this comparison (Fig.\ref{fig:benet16}) shows that the ice/water and water/vapor surfaces of the premelting film fluctuate independently and match exactly those corresponding to interfaces of the bulk phases up to wave-vectors of $q\approx 1.5$~nm$^{-1}$. This means that, for undulations of less than ca. 3~nm, these surfaces respond to  perturbations  exactly as the interfaces between bulk ice-water or water-vapor. The resemblance between surface fluctuations of premelting films and bulk water may be understood by comparison of the corresponding density profiles, which are almost the same at the interfacial regions (Fig.\ref{fig:benet16}-(c)). 

For fluctuations of larger scale, the bounding surfaces eventually start noticing the finite thickness of the premelting film and become highly correlated. Eventually, they must behave as one single fluctuating surface.\cite{chernov88} The distinction then can become dramatic dependent on the surface anisotropy.  The basal face is smooth, so that it effectively has an infinite stiffness to surface fluctuations. But all other faces of ice are rough at the triple point,\cite{baran24b} and will exhibit a compound surface tension very close to the sum of $\gamma_{\sl}+\gamma_{\lv}$ as discussed above.

\begin{figure*}[t]
	\centering
	\begin{minipage}[b]{0.5\textwidth}
		\begin{overpic}[width=\textwidth]{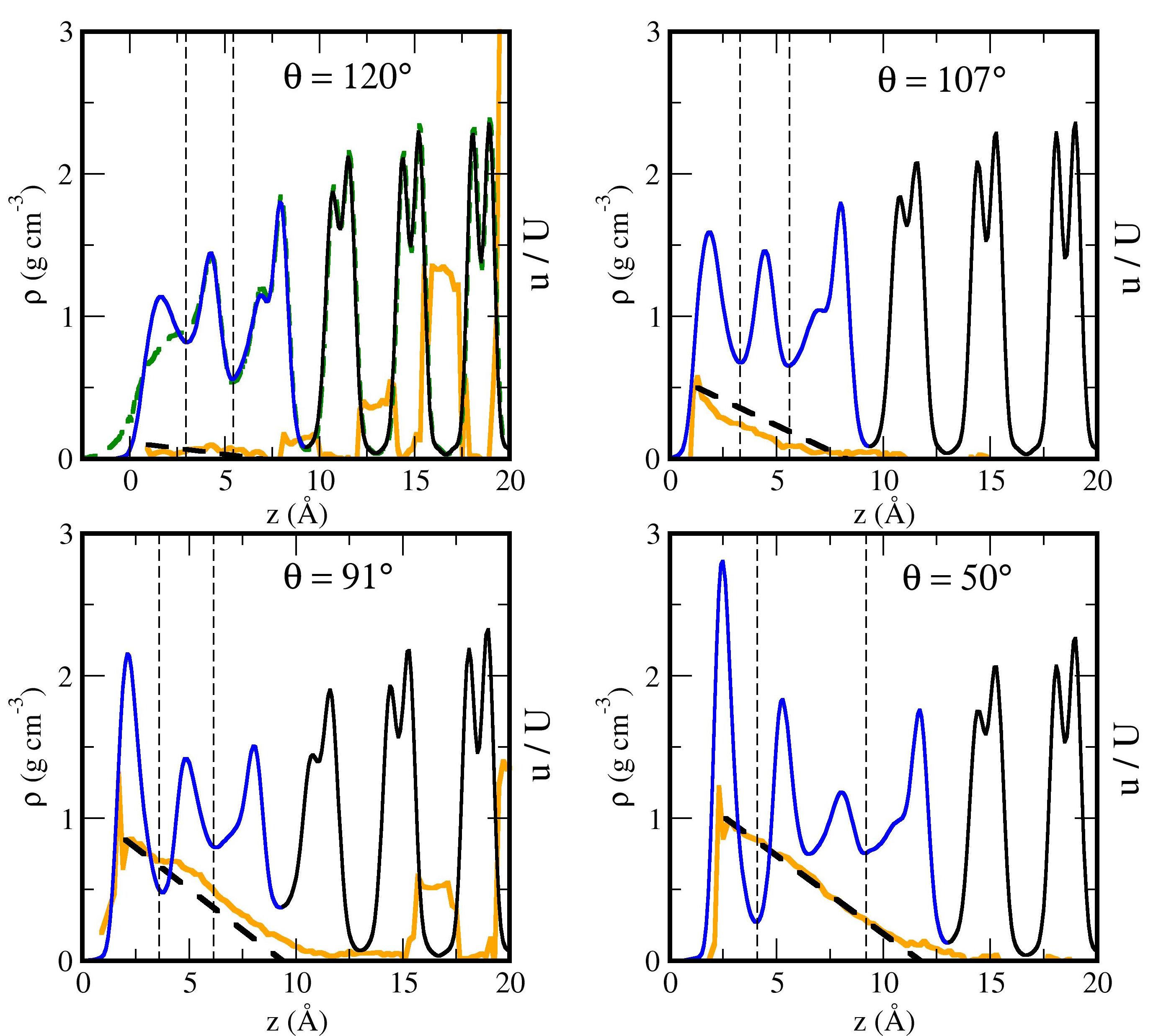}
			\put(15,92){\textbf{(a)}} 
		\end{overpic}
	\end{minipage}%
	\hfill
	\begin{minipage}[b]{0.5\textwidth}
		\begin{overpic}[width=\textwidth, trim=0 0 5cm 1cm,clip]{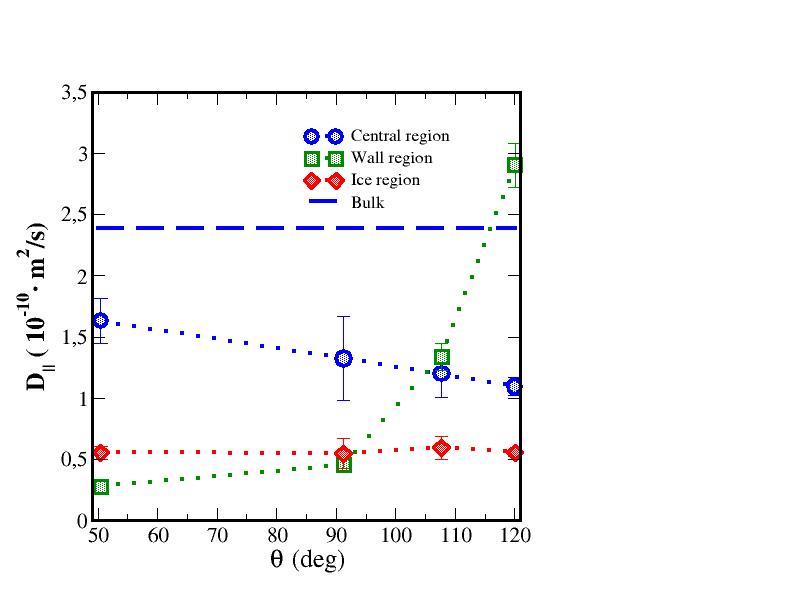}
			\put(15,92){\textbf{(b)}} 
		\end{overpic}
	\end{minipage}
	\caption{Rheology of premelting films at $T=-8$~C$^{\circ}$ as described
	   by the TIP4P/Ice model. Results are shown for \change{ice at the
	   interface with  four substrates with differing water contact angle}.
	   (a) Density profiles shown in either blue (for regions with liquid-like
	   molecules as the majority phase) or black (for regions with solid-like
	   molecules as the majority phase) continuous lines. The orange line
	   shows the corresponding flow profiles resulting when the substrate
	   slides at a speed of 5~m$/s$ on the ice surface. The dashed black lines show predictions for Couette flow, with slopes as given by the corresponding bulk water viscosity. (b) Diffusion coefficients calculated for the parallel motion of liquid-like molecules. The blue dashed line shows the result corresponding to bulk undercooled water. This is compared with parallel diffusion coefficients of water-like molecules in three different regions of the premelted layer as indicated by the dashed black lines of panel (a). Blue circles correspond to liquid-like molecules within the central region of the premelting film. Green squares correspond to molecules within the wall adsorption layer. The red diamonds correspond to the molecules in the ice adsorbed layer.  Reproduced with permission from  Baran et al.\cite{baran22} PNAS {\bf 119} e2209545119 (2022). 
	}
	\label{fig:flow}
\end{figure*}

The question then remains as to whether the small extent of liquid in between
the surfaces is sufficient to exhibit a dynamic response similar to bulk water.
Computer simulations are particularly insightful as they allow to monitor single
molecule motion without perturbing the system.  Unfortunately, measurements of
self-diffusion coefficients are difficult to interpret, as they strongly depend
on the location of molecules within the premelting
film.\cite{louden18,kling18,moreira18} The results typically show that molecules
close to the ice/water surface have a low diffusivity, while those at the
water/vapor
surface exhibit enhanced diffusion.\cite{louden18} However, bulk properties
cannot be expected for the three dimensional motion of the molecules, because of
the strong confinement  in the direction perpendicular to the film. Indeed, the
motion in this direction can even become sub-diffusive or glassy-like, because
of the locking and freezing of the molecules at the ice/water surface.\cite{moreira18}
On the other hand, measuring the diffusion parallel to the interface for
liquid-like molecules within the film, yields a rather different picture. In
this case, the water molecules exhibit self-diffusion coefficients that are very
close to those obtained for bulk undercooled water at the same temperature (Fig.\ref{fig:flow}). This
is sufficient for the film to display an almost bulk like rheological response
when sheared. Indeed, simulations of the TIP4P/Ice model displayed a velocity
pattern consistent with Couette flow, and a shear stress similar to that
expected for bulk water, albeit with slip lengths in the scale of several
nanometers  (Fig.\ref{fig:flow}).\cite{baran22} This results appear strongly
in conflict with experimental measurements of viscosity coefficients estimated
from the spreading of a liquid droplet siting on ice, which were found to be 20
to 200 times larger than for bulk water.\cite{murata15,nagata19,sazaki22}
However, based on the description of wetting dynamics of premelting films, the
motion of the contact line in a liquid drop is often dictated, not by the
spontaneous flow of the liquid film, but by the growth rate of the ice surface
underneath.\cite{sibley21} Therefore, the flow of a liquid film next to a
droplet need not be a faithful proxy of the fluid properties. Indeed, the current understanding of 
nanofluidics is that there is no reason to expect such large viscosities in films of nanometer 
thickness.\cite{bocquet10}

\section{Future Challenges}

\label{sec:concl}

The topic of ice premelting has been a puzzling problem of chemical physics,
with long standing debates over its mere existence, the order of magnitude of
the film thickness, its resemblance with the liquid phase and its significance
in important atmospheric processes such as snow flake growth. Unfortunately,
some of the conflicts result from interpretations which concentrate on one
single laboratories' piece of evidence, and disregard   other sources and experimental
methodologies. 
Using a wider perspective and a suitable
theoretical framework, many of the apparently conflicting observations are
actually found to be mutually consistent and complementary. 

In this perspective I have showed that blending a number of theoretical tools, including  wetting physics,\cite{llombart20,sibley21,luengo22b} intermolecular forces,\cite{luengo22b} statistical mechanics of crystal growth,\cite{benet16,benet19}  out of equilibrium physics,\cite{sibley21} and computer simulations,\cite{benet16,benet19,llombart19,llombart20,llombart20b,baran22,baran24b,baran25} a consistent picture is gradually emerging which allows to reconcile a large body of  apparently conflicting experimental results.  

In this view it appears that an equilibrium premelting film of sub-nanometer thickness gradually builds up atop pristine ice surfaces along the sublimation line,\cite{slater19,nagata19} and reaches a thickness of the order of one nanometer in the neighborhood of the triple point, whereupon, attractive van der Waals forces inhibit further growth in a case of incomplete surface melting.\cite{luengo22b} The nanometer  thickness is sufficient for the premelting liquid film to exhibit properties that are very similar to those expected for bulk liquid water.\cite{bocquet10,benet16,baran22} 

Whereas the equilibrium film thickness is bound by van der Waals intermolecular forces,\cite{luengo22b} 
diverging liquid films can emerge readily for small deviations away from equilibrium, when supersaturation is above a stability limit, as dictated by the crystal growth mechanism and the underlying intermolecular forces. In this situation,  the vapor condensation occurs at a faster pace than the freezing of the water film in contact with the solid surface, so that the liquid film gradually diverges in thickness. In this peculiar scenario, ice growth from the vapor proceeds in a wet mode, by freezing from the liquid film formed atop.\cite{sibley21}
This is a case of {\em kinetic} surface melting, which could easily be mistaken with {\em equilibrium} surface  melting, but is in fact an out of equilibrium phenomenon. 

Below the liquid film's stability limit, but above a kinetic condensation line, liquid droplets can condense from the vapor atop a marginally stable premelting film, leading to the appearance of inhomogeneities on the ice surface. The condensation of a droplet at water supersaturation or above is actually the expected behavior, and such droplets should not be confused with uniform equilibrium layers expected in equilibrium.

In between the sublimation line and  the kinetic condensation line, growth proceeds in a quasi-stationary mode, whereby accretion of ice occurs while a stable premelting film remains in a steady state regime of constant thickness.\cite{kuroda82,kuroda83,kuroda90} This premelting film, even at sub-nanometer length-scales that occur at strong under-cooling, has a profound impact on the crystal surface structure and promotes  a sequence of structural surface phase transitions,\cite{llombart20b} similar to those conjectured by Kuroda and Lacmann many years ago.\cite{kuroda82} As a result, growth rates of the basal and prism faces crossover twice, and lead to the alternation of columnar and plate like hexagonal prisms  described in the celebrated Nakaya diagram.\cite{nakaya54,furukawa07,libbrecht22}

In this account, a significant progress in the qualitative understanding of
equilibrium and out of equilibrium  premelting films appears to emerge. However,
there still remain important  problems to be understood, and a large number of properties to constrain within 
reasonable accuracy.

Indeed, to a great extent the conclusions drawn in this perspective rely on
computer simulations, which have been used both as a stand alone
characterization technique, or as an essential tool for the interpretation of
experiments.\cite{sanchez17,rashmi25,berrens25} However, simulations have some
limitations. The most apparent for the study of ice premelting close to the
triple point is the uncertainty in the determination of melting temperatures,
which due to system size effects and unavoidable simulation artifacts (finite
time step, truncation of interactions) is of the order of one
Kelvin.\cite{baran24b} On the other hand, at lower temperatures that are
relevant for the study of incipient surface disorder, classical molecular
dynamics could be unreliable at least in quantitative terms, as the strong
libration frequencies of water molecules result in significant nuclear quantum
effects.\cite{paesani08} Be as it may, most of the studies of ice premelting by
computer simulations have been performed with fixed point charge
models,\cite{abascal05,abascal05b} which do not deal with environment dependent
polarizability effects that could become very important at an interface. Another
problem is that most fixed charged models predict a very small vapor
pressure,\cite{vega06b} which could be a matter of concern. Current advances in force field modeling are promising,\cite{bore22} and preliminary assessments of the extent of ice premelting by means of accurate first principle models such as MDPol seem to confirm the most significant features observed with classical force fields.\cite{rashmi25} Qualitative agreement from machine learned Density Functional Potentials would be also welcome, and some work in this direction has been published recently.\cite{zeng24,berrens25} 
Another important issue \change{in} the study of ice premelting \change{with} computer simulations is the initial preparation of ice surfaces. Both basal and primary prism faces have a bilayer structure, and initial configurations can be prepared with a fully terminated or a half terminated bilayer. Some of the work reported in this study for the basal face was prepared with a half terminated bilayer,\cite{llombart19,llombart20,llombart20b} while that of the prism face was prepared with a fully terminated bilayer. Exactly what surface preparation is more relevant to the experimental situation is unclear, and further work in this direction is required to understand the actual self adsorption state of equilibrium premelting.

Most of this perspective article has been concerned with ice premelting above 200~K. However, the ice surface at lower temperatures is also of relevance for the understanding of atmospheric and extraterrestrial physics, and poses interesting fundamental problems concerning the  organization of dangling hydrogen bonds,\cite{nojima17}, the competition between cubic and hexagonal ice,\cite{hudait16} and surface reconstruction.\cite{fletcher92,hong24}  Some time ago, Fletcher suggested that the ice surface at low temperature could be stabilized by adapting ordered hydrogen bond arrangements, in a so called Fletcher stripe phase.\cite{fletcher92} This hypothesis has been explored in some studies for frozen realizations of bulk ice,\cite{buch08,pan10} but some work remains to understand how thermal agitation could change the picture, and what is the role of the conjectured  order/disorder transition in the emergence of translational surface disorder.\cite{sugimoto19} A particularly interesting question is how the disordering of the Fletcher surface phase could be coupled to the underlying bulk hydrogen bond network, and whether the Fletcher phase is consistent with the hydrogen bond disordered network of ice Ih.\cite{sugimoto19,berrens25} Recent experimental work shows that the low temperature surface reconstruction of epitaxially grown patches could be actually far more complicated \change{than} previously envisaged.\cite{hong24}

To great extent, of course, the progress that has been achieved rests on a large
body of experimental evidence gathered from a number of different techniques over the last few decades. In recent times, scrutiny of the interfacial structure using Sum Frequency Generation Spectroscopy (SFG) has received a great deal of attention, as it promises to disentangle bulk from surface properties better than other probes.\cite{wei01,bluhm02,sanchez17,smit17,nojima17} Unfortunately, the interpretation of the non-linear electronic polarizability, the difficulties of band assignment and the actual depth of the surface probe are highly controversial topics among expert practitioners which limit considerably the significance of the surface spectra so retrieved.\cite{yamaguchi19,shen20}  In this regard, current interpretation of the results relies to a great extent in the comparison of synthetic spectra obtained from computer simulations, and some promising advances in this direction have been reported very recently.\cite{rashmi25,berrens25}

However, the most insightful and clear
experiments appear to be those which combine in one or other way advanced
microscopy techniques, together with an exquisite control of the vapor
pressure.\cite{elbaum93,sazaki12} Without this, the entanglement between
freezing and condensation that is prone to occur inevitably as the triple point
is approached, leads to very difficult interpretation. A great progress could be
achieved if somehow the Brewster reflectometry used by Elbaum and collaborators
to measure film thickness,\cite{elbaum93} could be combined with the advanced
optical microscopy and vapor pressure control achieved by Sazaki and
collaborators.\cite{sazaki12,murata16} This would allow to characterize film
thicknesses of the pristine ice faces in equilibrium, as well as the
condensation effects that take place out of equilibrium. In this regard, it is
important to notice that most of the work of the Hokaido team reports results
close to crystal edges, which are expected to be indeed the most dynamically
active.\cite{frank82,libbrecht22} Yet, it is required to fully understand to
what extent the complex phenomenology they observe is related to general
features of the smooth mono-crystal surface, or to the proliferation of
dislocations and other defects caused from strain that often occur at the crystal edges.  The ice crystallites that are studied in that work are obtained from vapor deposition on Silver Iodide, which has a small but significant lattice mismatch. As far as understanding atmospheric crystal growth, this could pose problems, because the thickness of the resulting crystal is often not known, and the role of the underlying Silver Iodide mold in creating crystal defects is not fully understood. Ideally, the studies should be performed on perfect single monocrystals.\cite{shultz17,libbrecht22} A dream experiment would be the scanning of pristine hexagonal ice prisms grown from the vapor (diamond dust),\cite{libbrecht22} with a combination of Brewster angle reflectometry,\cite{elbaum93} and advanced optical microscopy,\cite{sazaki12} in exquisite vapor pressure control.\cite{murata16}

Be as it may, the current equipment available in Hokaido appears to be in a situation to confirm salient features of the kinetic phase diagram that emerges from the theoretical approach.  Particularly, the formation of complex crystal habits displayed in the Nakaya diagram occur somewhat above the equilibrium condensation line.\cite{harrington24} It is very tempting to speculate that this threshold of complex habit behavior is related to the
first emergence of water vapor condensation effects on the crystal surface as described by the Kinetic condensation line. However, the theoretical model shows that the steady state regime of crystal growth depends on the growth mechanism. Experiments have shown that, under favorable circumstances, either nucleated or spiral growth can be observed. Is it possible to detect the dependence of the kinetic condensation line on the prevailing  growth mechanisms?

 A significantly simpler experimental challenge would be to test the theoretical prediction of kinetic surface melting beyond the kinetic spinodal line.  In experiments, this would be detected as a temperature dependent threshold supersaturation where the ice surface becomes unstable and a strong condensation effect is observed throughout the whole crystal surface.  It is tempting to hypothesize that this regime could be related to hail formation. One way or the other, the confirmation of the most salient features of the kinetic phase diagram would be a major experimental achievement.

On the theoretical side, the best estimates of interface potentials clearly support a situation of incomplete wetting, with a shallow minimum of the interface potential at about one nanometer thickness that is consistent with the droplet contact angles obtained for vapor condensation.\cite{elbaum93,murata16} However, there are a number of important problems that are still difficult to explain from the model of surface intermolecular forces suggested here.

Firstly, on qualitative grounds, it is not clear  how to interpret the observation of a thick liquid film that appears to emerge below condensed liquid droplets at supersaturation.\cite{sazaki12,murata16} This phenomenology corresponds to the stabilization of a second minimum of the interface potential at intermediate film thickness.\cite{murata16} On theoretical grounds, this could result from either 1) oscillatory behavior of the short range contribution of the interface potential.\cite{llombart20}  2) the stabilization of a secondary minimum due to van der Waals forces.\cite{luengo22b} Both of these situations are possible in principle. However, the film observed in experiments is about 9~nm thick. This is too large to be the result of oscillations, and  too small to be the result of the sign reversal of the Hamaker constant.  So the problem rests to be solved. An important question here then is whether these thick film formed  is actually an equilibrium or a non-equilibrium phenomena. Indeed, both kinetic and thermodynamic interpretations of this observation may be found in the literature,\cite{asakawa16,murata16} and a full discussion on why should one or the other be favored is still lacking. If the thermodynamic interpretation were to be confirmed, then the only theoretical explanation that could be given within the current understanding of wetting and intermolecular forces is as follows: there is no secondary wetting state that could become an absolute minimum of the surface free energy. However, if a droplet is formed at supersaturation, and then local saturation at the contact line decreased transiently below the kinetic condensation line, a metastable minimum of the free energy would appear that could drive the formation of the observed film.   Confirmation of this scenario would require rather involved theoretical calculations of the diffusion limited evaporation/condensation dynamics in the neighborhood of the drop. 

Secondly, on quantitative grounds, the separation between the kinetic phase
lines of the kinetic phase diagram is given mainly by the depth of the interface
potential.\cite{sibley21}  However, the current estimate of this minimum
is less than one mJ$/m^2$, which   is extremely small.\cite{luengo22b,baran24b} Based on this calculation, 
the kinetic phase
lines displayed in Fig.\ref{fig:sazaki} should lie very close to each other and
differ by less than a Pascal.\cite{sibley21,luengo22b} The current experimental observations, however,
show lines that are separated by several many Pascal.\cite{murata16} The reason
for this discrepancy is not understood, but could be perhaps related to
diffusion limited growth of the droplets, a feature not accounted for in the
theoretical calculations.\cite{sibley21}

Another relevant conceptual problem refers to the role of electrostatic effects.
In the study of aqueous interfaces, the interface potential always features an
important contribution from electrostatic interactions. This have been discarded
altogether in the current survey, under the assumption that water is neutral.
However, in view of the shallow minimum of the interface potential, weak
electrostatic effects could become significant at long range. e.g. is there a
role of electrical effects in creating a second minimum of the interface
potential...? Currently, both experimental,\cite{dosch96,petrenko94} and
theoretical reports,\cite{watkins10,slater19} suggest that the first few ice
layers below the premelting film could exhibit significant Bjerrum defects,
leading to an effective surface charge at the ice/water surface. This, together
with charge accumulation from  water's self dissociation, might play a role in
the properties of premelting films,\cite{petrenko94} but an assessment with
simple force fields is simply beyond reach. The formulation of a simple
Poisson-Boltzmann theory is possible. But unfortunately, this requires imposing
boundary conditions at the ice/water and water/vapor interfaces that appear
currently not well understood. Indeed, the electrical nature of the water/vapor
interface is a highly controversial issue, and even the definition and sign of
the electric potential difference between the bulk phases appears to be
unsettled.\cite{becker22} Yet, a far less discussed topic that is needed to
interpret the role of electrostatics is the  boundary condition for the ionic
force at the solid phase: i.e. what is the pH of bulk ice? This was a problem
that attracted the attention of Eigen,\cite{eigen58} but does not seem to have
been a matter of inquiry in the literature ever since. Understanding both of these fundamental problems of chemical physics is interesting {\em per se}, and is absolutely necessary in order to assess whether electrostatic interactions promote or inhibit surface melting, and to what extent.

Regarding the shallowness of the interface potential, it is interesting to recall the discrepancy of contact angle measurements of macroscopic and microscopic droplets.\cite{huerre25} If the contact angle of 12$^{\circ}$ measured for large droplets were to be confirmed as the correct macroscopic equilibrium contact angle, the depth of the free energy minimum of the interface potential would increase significantly, and some of the problems mentioned above could be remedied. This begs the question: Could the discrepancy be related to line tension effects?
Unfortunately, assessing line tensions close to a wetting transition is
challenging, and even the sign of the line tension remains a controversial
issue.\cite{indekeu94} However, notice that assuming a negative line tension
would imply that the contact angle decreases as the droplets become        
smaller,\cite{navascues81,widom95} bringing into agreement the discrepancies between microscopic and macroscopic realizations of water droplets on ice. Yet, it is clear that the early stages of droplet deposition on the ice surface do not correspond to full thermodynamic equilibrium, so one cannot rule out that the discrepancy is related to pinning of the contact line by frozen water in some extremely complex out of equilibrium process.

This perspective has focused essentially on the study of pure ice samples. Yet, 
one of the most important motivations in the study of ice premelting is its role
in the scavenging of atmospheric gases.\cite{bartels-rausch13} Our current
understanding on the way trace gases and other impurities are adsorbed in the
premelting layer and partitioned between the bulk and interface remain rather
limited.\cite{wettlaufer99} Experiments and computer simulations have been
carried out recently on selected
adsorbents,\cite{bartels-rausch14,hudait17,llombart19,mitsui19,niblett21,ribeiro21,berrens22,richter25}  but a general understanding on this topic is still missing and requires further study.

Beyond surface ice premelting, it has been amply verified that liquid-like
layers can also be significant  at
interfaces,\cite{dash99,engemann04,liljeblad17} and some recent computer
simulations confirm this for a wide range of
substrates.\cite{ronneberg20,baran22,cui24,atila25} However, there is still a
large room for understanding the role of a substrate on ice premelting, and what
interfacial characteristics enhance or inhibit growth of quasi-liquid
layers.\cite{baran25} These interfacially premelted layers apparently exhibit
close to bulk like properties,\cite{baran22} and can provide for a lubrication effect in ice
sliding.\cite{baran22,baran24}


In summary, a significant progress in the understanding of ice premelting has been gathered in the last decade, but there remains a great room for further research. A promising but formidable challenge could be to combine microscopic insight towards the formulation of a crystal growth law (Figure \ref{fig:stepfree}) with advances in the mesoscopic modeling of crystal habits.\change{\cite{yokoyama90,barret12,demange17b,libbrecht22}} This could address a major task pending since first tackled by Kepler in 1611. \cite{kepler11} Can we possibly predict ab-initio the shape of snow crystals in the atmosphere from the underlying principles of physics?

\begin{acknowledgments}
	I would like to express my most sincere gratitude to my collaborators and
friends  who have helped me understand some of the intricate physics of ice
premelting.  First of all I am particularly indebted to former  PhD students and
PostDocs Jorge Benet, Pablo Llombart, Ram\'on Bergua, Juan Luengo, Fernando Izquierdo-Ruiz and
Lukasz Baran, 
who have performed  all of the computational work, and stimulated
me with many illuminating discussions, and particularly, to Lukasz Baran, for
invaluable discussions, help and ongoing collaboration. I am also greatly
indebted to my  colleagues, Eva G. Noya, David Sibley, Andrew Archer,
Mathias Bostr\"om and Eduardo Sanz for their willingness to embark in our collaboration and for many helpful discussions. 
	I also wish to express my gratitude for discussions and support  to
Valeria Molinero, Keneth Libbrecht, Kenichiro Murata, Dan Gezelter, Davide
Donnadio, Gilles Demange, Cristophe Josserand, Menno Demmenie, Emily Aesenath-Smith and many others. 
	The research discussed here was partially funded
	by the Agencia Estatal de Investigaci\'on (Ministerio de Ciencia y
	Econom{\'i}a) and Fondo Europeo de Desarrollo Regional (FEDER) under grant PID2023-151751NB-I00.   
\end{acknowledgments}


%
	
\end{document}